%% file: main.tex
\documentclass[11pt]{article}
\usepackage[a4paper,margin=1in]{geometry}
\usepackage{amsmath,amssymb,amsthm,mathtools}
\usepackage{microtype,booktabs,xcolor}
\usepackage{cite}
\usepackage{tabularx}
\usepackage{algorithm}
\usepackage[noend]{algpseudocode}
\usepackage{tikz}
\usetikzlibrary{arrows.meta,positioning,calc,decorations.pathreplacing}
\usepackage{authblk}
\usepackage[colorlinks=true,linkcolor=blue!55!black,citecolor=blue!55!black,urlcolor=blue!55!black]{hyperref}

\newtheorem{theorem}{Theorem}[section]
\newtheorem{lemma}[theorem]{Lemma}
\newtheorem{proposition}[theorem]{Proposition}
\newtheorem{corollary}[theorem]{Corollary}

\theoremstyle{definition}

\newcommand{\E}{\mathbb E}
\newcommand{\OPT}{\operatorname{OPT}}
\newcommand{\ALG}{\operatorname{ALG}}
\newcommand{\RoE}{\operatorname{RoE}}
\newcommand{\pos}[1]{\left(#1\right)_+}
\newcommand{\cost}{\operatorname{cost}}

\input{paper-style}
\title{Online Network Design with Maximum Delay\\and Stochastic Arrivals}
\author{Tianhang Lu}
\affil{School of Mathematical Sciences, Ocean University of China,\\
Qingdao 266100, Shandong, China,\\
lutianhang@stu.ouc.edu.cn}
\date{}

\begin{document}
\maketitle
\input{sections/abstract}
\input{sections/introduction}
\input{sections/model}
\input{sections/shared_overview}
\input{sections/facility_overview}
\input{sections/technical_tools}
\input{modules/ag/sections/maximum}
\input{sections/tree_applications}

\input{sections/facility_proof}

\input{sections/ai_declaration}
\input{references}
\end{document}

%% file: paper-style.tex
\usepackage{tcolorbox}
\tcbuselibrary{skins,breakable}
\usepackage{caption}
\definecolor{PaperBlue}{HTML}{315B88}
\definecolor{PaperTeal}{HTML}{21776F}
\definecolor{PaperAmber}{HTML}{A46422}
\definecolor{PaperPurple}{HTML}{705A87}
\definecolor{PaperInk}{HTML}{243449}
\hypersetup{linkcolor=PaperBlue,citecolor=PaperTeal,urlcolor=PaperBlue}
\tcbset{paperstatement/.style={enhanced,sharp corners,
  boxrule=.45pt,left=5pt,right=5pt,top=3pt,bottom=3pt,
  before skip=8pt,after skip=8pt}}
\tcolorboxenvironment{theorem}{paperstatement,
  colback=PaperTeal!5,colframe=PaperTeal!75!black}
\tcolorboxenvironment{lemma}{paperstatement,frame hidden,
  borderline west={1.5pt}{0pt}{PaperBlue},colback=PaperBlue!3}
\tcolorboxenvironment{proposition}{paperstatement,frame hidden,
  borderline west={1.5pt}{0pt}{PaperPurple},colback=PaperPurple!4}
\tcolorboxenvironment{corollary}{paperstatement,
  colback=PaperTeal!3,colframe=PaperTeal!50}
\tcolorboxenvironment{definition}{paperstatement,frame hidden,
  borderline west={1.5pt}{0pt}{PaperAmber},colback=PaperAmber!4}

%% file: sections/abstract.tex
\begin{abstract}
We study online network services that combine pending requests into
batches. Each batch pays for its spatial service and for the waiting
time of its oldest request. The objective is the sum of these costs
over all batches. Thus requests can share both a spatial purchase and
one maximum-delay charge. We assume independent Poisson arrivals of
known rates, while actual requests are revealed only upon arrival.

For stateless services, suppose that each type has an attainable
minimum cost when served alone, that every batch costs at least the
stand-alone minimum of each type it contains, and that one finite-cost
action can serve any batch, regardless of its request multiplicities. We give a
deterministic algorithm with long-run ratio of expectations $e/(e-1)$
against an unrestricted offline optimum. This factor is optimal,
even allowing randomization, already at one request location. The
key comparison measures the union of time intervals associated with
requests. The theorem applies directly to multi-root and pair
aggregation on trees. We also study facility location, where each
request pays its own connection distance and the bounded-action
assumption can fail. A deterministic policy gives a ratio $3$
for arbitrary nonnegative opening costs. It immediately serves a finite
initial set of requests, computes an assignment from their observed
arrival gaps, and then keeps that assignment fixed. Its analysis combines
Poisson insertion with a time-slice packing bound on the offline optimum.
Every guarantee includes the full cost of
completing finite inputs without an end-of-input signal.
\end{abstract}

%% file: sections/introduction.tex
\section{Introduction}\label{sec:intro}
Consider a network operator collecting requests for a shared service.
Opening a connection or activating a depot has a fixed cost, so the
operator would like several requests to use the same service. Yet a
batch cannot be postponed indefinitely: its oldest request has already
been waiting, and determines how late the batch is. A younger request
can join this batch without increasing its maximum waiting time.
This gives two reasons to combine requests: they share both spatial
resources and a single delay charge.

We study this tradeoff when each service pays its spatial cost plus
the maximum waiting time of its requests. The objective is the sum
of these costs over all services. This is a \emph{per-batch} maximum:
it is neither a deadline constraint nor the maximum wait over the
entire execution. For example, requests can be aggregated to one of
several roots in a tree, or pairs of endpoints can share connecting
edges. A facility can also serve a batch, paying an opening cost and
a separate connection distance for each request. In each case the
algorithm must decide both when to act and how to serve the batch.

The corresponding additive-delay problems, in which every request
pays its own wait, have been studied extensively over the last decade
\cite{bienkowski-et-al-2016,buchbinder-2017,azar-ganesh-ge-panigrahi-2017,
carrasco-2018,azar-touitou-2019,azar-touitou-2020,azar-setcover-2020,
touitou-lower-2021,mcm-2021,bienkowski-et-al-2022,le-2023,touitou-2023,
touitou-nonclair-2023,touitou-service-2025,ahmadian-2026}.
The frameworks of Azar and Touitou~\cite{azar-touitou-2019,azar-touitou-2020}
show that several spatial problems can be understood through a common
temporal method. Maximum delay changes that temporal method. Splitting
a batch preserves its sum of individual waiting times, but can repeat
its maximum-wait charge. Conversely, adding more requests to a batch
need not increase that charge at all.

Recent work already gives constant worst-case guarantees for maximum
delay. Lu et al.~\cite{lu-mla-2026} obtain the optimal deterministic
factor $2$ and randomized factor $e/(e-1)$ for rooted aggregation and
extend their results to submodular service costs. They also give
constant guarantees for mobile service with per-batch maximum
delay~\cite{lu-osd-2026}. Thus our motivation is not the absence of
constant algorithms throughout the maximum-delay setting. Instead,
we ask what statistical information reveals about the cost of batching,
and whether that information permits a sharp guarantee under weaker
assumptions on the spatial service.

In a repeated service system, historical demand may indicate how often
each type of request occurs. We model this information by independent
Poisson streams of known rates. A rate does not reveal an arrival time,
but the probability of seeing no request during a given interval is
known. This simple information turns out to be particularly useful for
maximum delay. It suggests the following question.
\begin{quote}\itshape
Can arrival rates identify the unavoidable long-run cost of shared
service, and can a simple causal schedule attain a sharp constant
factor against an optimum that knows all realized arrivals?
\end{quote}

For additive delay, Mari, Paw\l{}owski, Ren, and Sankowski reported
constant expectation ratios for stochastic matching~\cite{mari-et-al-2023}
and rooted aggregation~\cite{mari-et-al-2024}. Their work shows how
rates can guide a choice between immediate and periodic service.
Maximum delay calls for a different comparison with the optimum:
many waiting requests in one batch supply only one delay payment.
We show that the appropriate lower-bound object is the measure of a
union of time intervals. This observation leads to a sharp theorem
for stateless shared services and, with a separate spatial argument,
a constant guarantee for facility location.

\subsection{Our contributions}
We compare expected online and offline costs over increasingly long
horizons on a fixed finite instance. Algorithms know the rates, observe
requests as they arrive, and complete every finite input without being
told when arrivals stop. The offline optimum may choose its batches
and spatial services after seeing the entire input.
\begin{itemize}
\item \textbf{An optimal guarantee for shared services.}
We give a deterministic algorithm with ratio $e/(e-1)\approx1.582$
for a broad class of stateless services. Each type must have an
attainable minimum singleton cost that lower-bounds any batch serving
it. There must also be one finite-cost action that can serve all
pending requests, regardless of how many have accumulated. No tree
structure or submodularity assumption is required.

The factor is optimal over this class, even for randomized algorithms:
one request location already gives a matching lower bound. Our policy
also gives a finite maximum wait on each fixed instance. The result
shows exactly how far known arrival rates can close the gap between
online and offline decisions in this class.
\item \textbf{Consequences for tree aggregation.}
Multi-root and pair aggregation on trees both satisfy these conditions
and inherit the factor $e/(e-1)$. The offline optimum remains free to
choose a different root for each batch in multi-root aggregation.
These are applications of the shared-service theorem, rather than
separate temporal algorithms.
\item \textbf{Facility location with individual connections.}
For arbitrary finite metrics and nonuniform opening costs, we obtain
a deterministic algorithm with ratio $3$.
It immediately serves the first $m$ requests, where $m$ is the number
of positive-rate types, and uses their observed arrival gaps to compute
an assignment that remains fixed thereafter. No internal randomness is used.
The offline optimum may send different requests at the same location
to different facilities. This result requires a separate comparison:
connection costs grow with the number of requests, so a universal
spatial action of bounded cost need not exist.
\end{itemize}

\begin{table}[t]
\centering\small
\renewcommand{\arraystretch}{1.2}
\begin{tabularx}{\textwidth}{@{}Xc@{}}
\toprule
\textbf{Service model} & \textbf{Long-run ratio bound}\\
\midrule
Stateless shared service & $e/(e-1)$, optimal over the class\\
Multi-root and pair aggregation on trees & $e/(e-1)$\\
Facility location in finite metrics & $3$\\
\bottomrule
\end{tabularx}
\caption{Deterministic guarantees under known-rate Poisson arrivals,
against unrestricted offline schedules. The facility policy fixes its
assignment after a finite initialization phase. Expectations are over
input arrivals, and all costs of finite-input completion are included.}
\label{tab:results}
\end{table}

These are long-run guarantees. The shared-service policy has a fixed
maximum wait depending on the instance. The facility policy attaining
the stated factor uses increasing service intervals; fixed finite
intervals give factor $3(1+\varepsilon)$ for any $\varepsilon>0$
(Corollary~\ref{fl:max:cor:bounded-wait}). The finite
horizon corrections are established in the proofs. We do not claim
a uniform multiplicative bound for every short horizon.

For path-union costs, the same numerical factor also occurs in the
adversarial randomized result of Lu et al.~\cite{lu-mla-2026}. Here it
is attained deterministically using arrival rates. The broader theorem
requires only its stated singleton and universal-action conditions;
in particular, the multi-root application does not rely on a claim
that minimizing over roots preserves submodularity.

\subsection{Our techniques}
\paragraph{Measure shared delay by a union of time intervals.}
Each request draws an interval starting at its arrival, with length
equal to its cheapest individual service cost. All intervals belonging
to one offline batch fit inside an interval whose length is paid by
that batch's spatial cost and maximum wait. Taking the union of request
intervals therefore gives a lower bound on every offline schedule.
Unlike adding their lengths, taking a union does not charge a batch's
shared delay repeatedly. Poisson rates let us compute how much of the
time line is covered in expectation.

\paragraph{Use rates to choose between immediate and shared service.}
For light demand, serving requests individually is inexpensive. For
heavy demand, a long enough collection period spreads the cost of a
universal service over time. Adding requests to a batch does not add
individual waiting charges, which makes this second option effective.
Comparing these two choices with the coverage lower bound gives the
factor $e/(e-1)$. A one-location example shows why no online policy
can improve the universal factor.

\paragraph{Handle individual connection costs separately.}
For facility location, batching cannot remove the distance charged to
each request. We immediately serve a finite initial set of requests
and scale their observed arrival gaps to obtain independent exponential
variables with the required rates. These values determine balls whose
sets of contained facilities are greedily packed.
Poisson insertion bounds the expected assignment cost by three times
an auxiliary packing count. We apply the same packing rule at each
time slice to shrinking balls associated with actual requests.
Within one offline batch, at most one accepted ball contains its
facility at any time. The opening and maximum wait pay for these
portions; connections pay for the remaining portions.
Stationarity and offline subadditivity turn this certificate into
the lower bound needed to justify the assignment. The future arrival
stream is independent of the initialization history, so the same
comparison applies to the subsequent service phase. Initialization
adds only an instance-dependent constant to expected cost.

\input{figures/framework}

\subsection{Related work}
\paragraph{Additive aggregation and network design.}
Bienkowski et al.~\cite{bienkowski-et-al-2016} initiated the study
of aggregation on trees of arbitrary fixed depth with general delay.
The deadline bounds were improved to $O(D)$ by Buchbinder
et al.~\cite{buchbinder-2017} and to $D$ by McMahan~\cite{mcm-2021}.
For general delay, Azar and Touitou~\cite{azar-touitou-2019}
obtained $O(D^2)$; the recent preprint of Ahmadian
et al.~\cite{ahmadian-2026} gives $2D$.
Carrasco et al.~\cite{carrasco-2018} study a general set aggregation
model with logarithmic dependence on the number of requests.
Azar and Touitou's later framework~\cite{azar-touitou-2020}
reduces network design with deadlines or delay to offline optimization
problems with an additional logarithmic loss. Set cover with delay
admits polylogarithmic guarantees even without knowing future delay
functions~\cite{azar-setcover-2020}, and nearly matching lower bounds
are known for set cover and several network design extensions
\cite{touitou-lower-2021}. These objectives sum requests' delays;
the maximum-delay accounting used here is different.

\paragraph{Facility location and online service.}
The adversarial facility-location results of Azar and Touitou
\cite{azar-touitou-2019,azar-touitou-2020} are complemented by the
linear-delay algorithms of Bienkowski et al.~\cite{bienkowski-et-al-2022}.
The latter distinguish a one-sided model with an
$O(\log m/\log\log m)$ guarantee from a two-sided model with a
constant guarantee. Neither objective is the per-batch maximum used
here. Azar et al.~\cite{azar-ganesh-ge-panigrahi-2017} introduced
mobile service with delay. The general-metric guarantee became
randomized $O(\log^2 n)$ in~\cite{azar-touitou-2019} and deterministic
$O(\log n)$ in~\cite{touitou-2023}. A persistent mobile server has
state-dependent spatial costs and lies outside our stateless theorem.

\paragraph{Clairvoyance and stochastic information.}
Le, Umboh, and Xie~\cite{le-2023} distinguish knowing future delay
functions from observing only their current values. Touitou develops
nonclairvoyant network design frameworks~\cite{touitou-nonclair-2023}
and a nearly optimal nonclairvoyant service
algorithm~\cite{touitou-service-2025}. Our delay rule is known;
the uncertainty concerns future arrivals. Mari
et al.~\cite{mari-et-al-2023,mari-et-al-2024} study known Poisson
rates and report constant expectation ratios for additive-delay matching
and rooted aggregation. We adopt this stochastic perspective, while
replacing individual-wait accounting by interval coverage. Rates are
assumed known and are not estimated by the algorithms in this paper.

\paragraph{Maximum delay.}
Lu et al.~\cite{lu-mla-2026} give an exact polynomial-time offline
algorithm and tight deterministic and randomized online factors for
rooted aggregation, extending these statements to submodular service
costs. Their mobile-service work~\cite{lu-osd-2026} gives deterministic
factors $10$ on lines, $12$ on trees, and $20$ on arbitrary finite
metrics. These recent preprints demonstrate that maximum delay can
admit constants already under adversarial arrivals. We study how
stationary arrival information yields a sharp deterministic factor
for a broader stateless class and a deterministic facility policy whose
assignment is fixed after initialization.
Our guarantees are asymptotic expectation ratios, not improvements
to those worst-case competitive bounds. In particular, the calendar's
instance-dependent completion term cannot be removed to obtain a
uniform bound at every finite horizon.

\subsection{Organization}
Section~\ref{sec:model} defines the model and basic Poisson facts.
Section~\ref{sec:shared-story} develops the shared-service idea from
one location through its tree applications. Section~\ref{sec:facility-story}
explains the separate facility argument. These sections give the main
results and the comparisons behind their constants.
Section~\ref{sec:technical-tools} begins the technical development.
Sections~\ref{ag:sec:maximum} and~\ref{ag:shared:sec:applications}
give the complete shared-service proof and tree verifications;
Section~\ref{sec:facility} proves the facility guarantee.

%% file: figures/framework.tex
\begin{figure}[t]
\centering
\begin{tikzpicture}[font=\small,>=Latex,
  box/.style={draw,rounded corners=2pt,align=center,text width=4.5cm,
    minimum height=1.05cm,inner sep=6pt}]
\node[box,draw=PaperAmber,fill=PaperAmber!5] (interval) at (0,0)
  {Singleton intervals\\$\E[\OPT_T]\ge(1-e^{-L})(T-s_*)_+$};
\node[box,draw=PaperTeal,fill=PaperTeal!5] (calendar) at (0,-2.0)
  {Bounded universal action\\immediate service or fixed calendar};
\node[box,draw=PaperTeal,fill=PaperTeal!5] (trees) at (0,-4.0)
  {Multi-root and pair aggregation\\sharp universal factor $e/(e-1)$};
\draw[->,PaperTeal,thick] (interval)--(calendar);
\draw[->,PaperTeal,thick] (calendar)--(trees);
\node[box,draw=PaperBlue,fill=PaperBlue!4] (facility) at (6.4,0)
  {Facility connections\\cost grows with multiplicity};
\node[box,draw=PaperAmber,fill=PaperAmber!5] (packing) at (6.4,-2.0)
  {Poisson balls\\and time-slice packing\\assignment certificate};
\node[box,draw=PaperBlue,fill=PaperBlue!4] (assignment) at (6.4,-4.0)
  {Observed arrivals\\fixed assignment\\deterministic factor $3$};
\draw[->,PaperBlue,thick] (facility)--(packing);
\draw[->,PaperBlue,thick] (packing)--(assignment);
\end{tikzpicture}
\caption{Two uses of temporal coverage. A bounded universal action
completes the shared-service argument (left). Per-request connections
require a separate spatial assignment and packing argument (right).
The facility assignment is computed from a finite observed prefix.}
\label{fig:framework}
\end{figure}
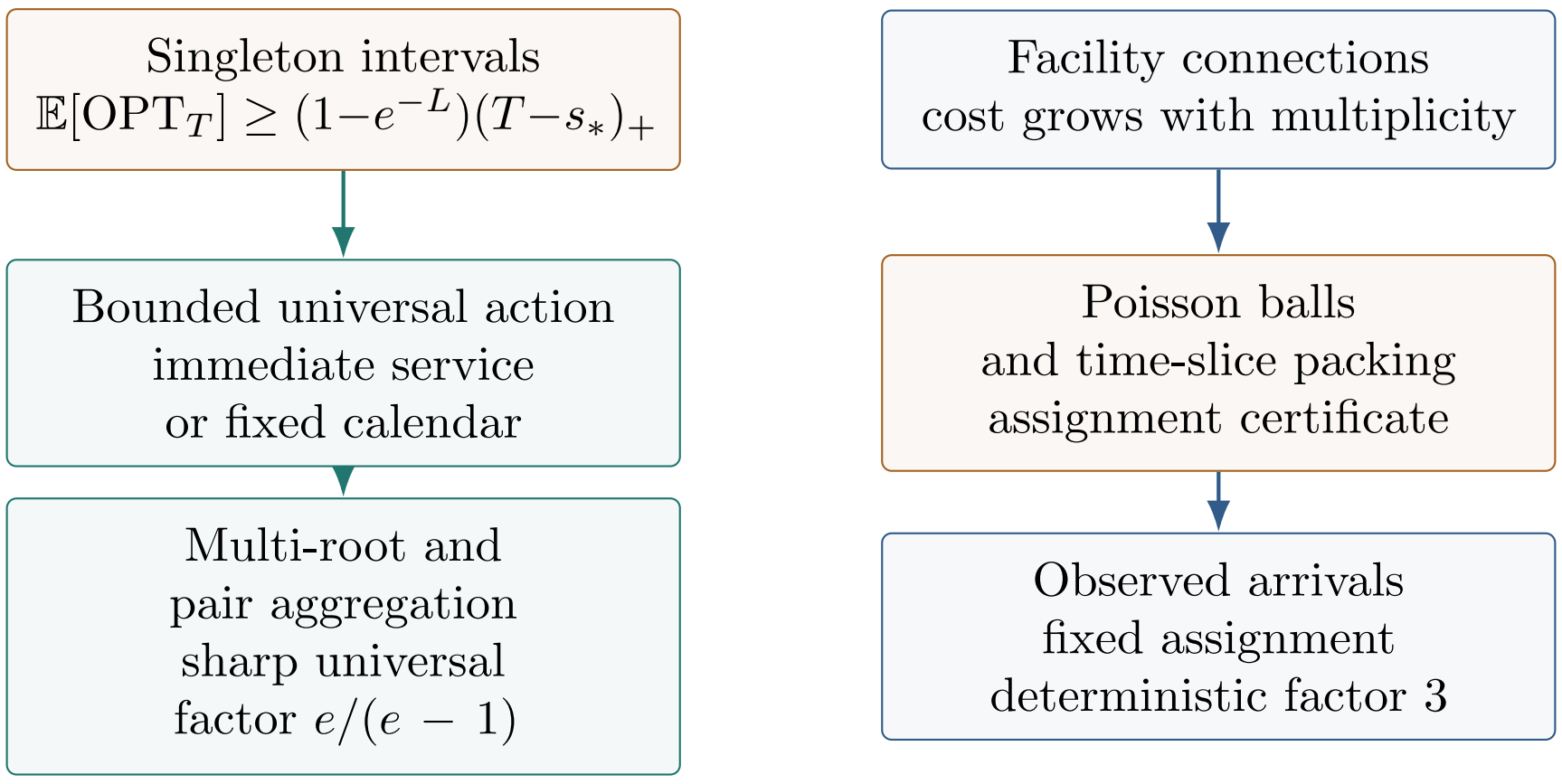

%% file: sections/model.tex
\section{Model and basic Poisson facts}\label{sec:model}
\subsection{Requests and maximum-delay batches}
There are finitely many request types. Each occurrence $i$ has a type
$x_i$, an arrival time $a_i$, and a service time $t_i\ge a_i$.
A service chooses a nonempty batch $B$ of pending requests and a
feasible spatial action. If its spatial cost is $C(B)$ and it occurs
at time $t$, the total batch cost is
\[
 C(B)+\max_{i\in B}(t-a_i)=C(B)+t-\min_{i\in B}a_i.
\]
The objective sums these costs over services. Maximum delay means
one charge for the oldest request in each batch. It is neither a
single maximum over the whole execution nor a deadline constraint.
Simultaneous services pay their own maximum delays separately.

Services are instantaneous and have no capacity restriction. They
are stateless: buying a network or opening a facility serves the
current batch, without retaining a purchased network or an open
facility for the next batch. The general shared-service theorem
will specify two conditions on the available spatial actions.
It does not cover a persistent mobile server.

For tree aggregation, a service buys the edges connecting its requests
to a chosen root, or connecting the endpoints of pair requests. For
facility location in a finite metric, service at $f$ pays an opening
cost $c_f\ge0$ and the distance $d(x_i,f)$ for each occurrence $i$.
A later service pays for its spatial action again. The precise
feasibility conditions are stated when each model is considered.

\subsection{Poisson arrivals and the comparison}
Type $x$ follows an independent homogeneous Poisson process of known
rate $\lambda_x\ge0$. Its positive-rate interarrival times are
independent exponentials of mean $1/\lambda_x$. The algorithm knows
the rates and spatial instance, but sees only arrivals that have
already occurred. Zero-rate types can be ignored. Arrivals at a
service time are observed first, and simultaneous actions are
processed in execution order.

Let $I_T$ be the arrivals in $[0,T]$. The algorithm receives no signal
that the input has ended and must eventually serve every request.
Its cost $\ALG_T=\ALG(I_T)$ includes all purchases and waiting after
$T$ needed for completion. The offline optimum $\OPT_T$ knows the
entire realized input and is not restricted to the algorithm's
assignments or calendar. For each fixed instance with positive
optimal asymptotic cost, our criterion is
\begin{equation}\label{eq:roe}
 \RoE(\ALG)=\limsup_{T\to\infty}
 \frac{\E[\ALG_T]}{\E[\OPT_T]}.
\end{equation}
Expectations include independent algorithmic randomness when relevant.
This is a ratio of expected total costs, rather than an expectation
of samplewise ratios. The notation $O_I(1)$ allows dependence on the
fixed instance, but not on $T$. An asymptotic guarantee does not remove
the need to pay for completing the last batch.

\subsection{The probability facts behind the coverage argument}
A fixed interval of length $s$ contains $\lambda s$ arrivals in
expectation and is empty with probability $e^{-\lambda s}$. Counts
in disjoint intervals are independent, as are counts from different
type streams. Combining independent streams adds their rates.

These facts let us compute the probability that a fixed point on the
time line is covered by a request interval. If type $x$ covers that
point whenever it arrives in a prescribed interval of length $s_x$,
the probability that no type covers it is
\[
 \prod_x e^{-\lambda_xs_x}=e^{-\sum_x\lambda_xs_x}.
\]
The intervals here are fixed independently of the arrivals. We never
assume the same calculation holds for a batch chosen by the offline
optimum after observing the input. The more general counting
identities used in the facility and sharpness proofs are developed
in Section~\ref{sec:technical-tools}.

%% file: sections/shared_overview.tex
\section{Shared service: from one location to a general theorem}
\label{sec:shared-story}
We begin with one request location. This example explains both the
algorithm's timing choice and an exact description of offline cost.
The lower-bound part of that description then extends to many types
without requiring a tree or a submodular spatial cost.

\subsection{Why longer batches behave differently}
Let requests arrive at rate $\lambda$ and let one service cost $c$,
independently of the number of requests it serves. Immediate service
has expected cost rate $\lambda c$. If we instead clear the queue
every $H$ time units, a nonempty batch costs at most $c+H$.
Its long-run cost rate is therefore at most
\[
 1+\frac cH.
\]
A longer period spreads the purchase over more time, while the maximum
delay contributes at most one unit per unit time. Unlike additive
delay, more arrivals do not create additional waiting charges inside
the same batch. This suggests choosing between immediate service and
large shared batches, with benchmark rate $\min\{\lambda c,1\}$.
It remains to compare this choice with an optimum that sees the future.

\subsection{Draw one interval per request}
For every request arriving at $a$, draw the interval $[a,a+c]$.
Suppose a batch is served at time $t$, and its first request arrived
at $a_-$. All its request intervals lie inside $[a_-,t+c]$.
That envelope has length $c+t-a_-$, exactly the cost of the batch.
Taking the union of request intervals thus gives a lower bound on
the cost of any schedule.

At one location this lower bound is exact. Group consecutive arrivals
whose intervals overlap, and serve each group at its last arrival.
The cost is precisely the length of that connected interval component.
For example, with $c=1$ and arrivals at $0$, $0.4$, and $2$, the union
is $[0,1.4]\cup[2,3]$. Its length is $2.4$, attained by serving the
first two requests at $0.4$ and the last at $2$. Serving all three
together would cost $3$.

This representation converts offline optimization into a coverage
calculation. Away from the beginning and end of the input horizon,
a point is uncovered precisely when the preceding interval of length
$c$ contains no arrival. Hence
\[
 \lim_{T\to\infty}\frac{\E[\OPT_T]}T=1-e^{-\lambda c}.
\]
The key is that these request intervals are fixed by arrivals and
singleton costs. We do not try to apply a Poisson formula to intervals
chosen by the offline schedule.

\subsection{Which properties of a spatial service do we need?}
For each type $x$, let $s_x$ be an attainable minimum spatial cost
of serving one request alone. We require every batch serving $x$ to
have spatial cost at least $s_x$. We also require one fixed action
that serves any multiset of pending requests for spatial cost at most
some finite $M$, independent of multiplicities. These conditions concern
stateless services; all spatial costs are paid anew at each event.

Now draw $[a_i,a_i+s_{x_i}]$ for request $i$. If its batch has spatial
cost $C$, every interval in that batch lies in $[a_-,t+C]$. The same
envelope argument proves
\[
 \OPT_T\ge\left|\bigcup_{i\in I_T}[a_i,a_i+s_{x_i}]\right|.
\]
Unlike the one-location case, equality is not needed or asserted.
The singleton costs supply the lower bound; the universal action
will supply an online upper bound.

Write $L=\sum_x\lambda_xs_x$, the cost rate of immediate singleton
service. Independence gives coverage probability $1-e^{-L}$ away
from the initial boundary. Thus the lower bound has long-run rate
$p=1-e^{-L}$. It counts the union of time intervals, not their sum,
so a batch's single maximum-delay charge is never used repeatedly.

\subsection{The algorithm and the constant}
Let $\rho=e/(e-1)$. The algorithm makes one decision from the known
instance before arrivals begin.
\begin{itemize}
\item If $L\le1$, serve each request immediately by a cheapest
singleton action. The expected rate is $L$, and
$L/(1-e^{-L})\le\rho$ for $0<L\le1$.
\item If $L>1$, use the universal action at regular times, skipping
empty queues. A period $H$ gives cost rate at most $1+M/H$.
Since $\rho(1-e^{-L})>1$, a finite period can make this rate equal
to the target $\rho(1-e^{-L})$.
\end{itemize}
Explicitly, the second choice is
\[
 H=\frac{M}{\rho(1-e^{-L})-1}.
\]
At $L=1$ the denominator would vanish, but immediate service already
achieves the target there. Thus every instance uses either immediate
service or a finite fixed period. Types with free singleton service
are served immediately; if $L=0$ all requests are free.

The result is a deterministic long-run ratio at most $e/(e-1)$.
Every finite input finishes by the next scheduled clearing. The
complete bound has an additive term depending on the instance, and
an expensive universal action or a load just above one may yield a
long period. Theorem~\ref{ag:shared:thm:shared} gives the finite-horizon
bound; Section~\ref{ag:shared:sec:finite} explains why
this policy need not have a uniform short-horizon ratio.

\subsection{Why the factor is optimal}
Return to one location with $\lambda c=1$. The offline rate is
$1-e^{-1}$. Any online policy must instead pay at rate at least one.
The reason is a balance between an empty and a nonempty queue.
While the queue is empty, a new arrival starts a batch that eventually
needs a purchase. The expected rate of this purchase obligation is
$\lambda c=1$. While the queue is nonempty, its oldest request
accumulates waiting at unit rate. One of these two accounts is always
active.

The formal proof first shows that clearing the whole queue can only
help, then makes this accounting precise, including a batch unfinished
at the horizon. It applies to randomized as well as deterministic
causal policies. Therefore no universal factor below
\[
 \frac1{1-e^{-1}}=\frac e{e-1}
\]
is possible. The complete sharpness argument follows the upper bound
in Section~\ref{ag:sec:maximum}.

\subsection{Two tree applications}
For multi-root aggregation, one request's singleton cost is the
cheapest root activation plus the path from that root to the client.
One root and a subtree containing all possible clients provide a
universal action. These two observations are enough: the offline
optimum can still choose a different root for every batch.

For pair aggregation, a type's singleton cost is the weight of its
required endpoint path. A connected subtree containing all possible
endpoints is a universal action. If a disconnected path union is
allowed as one service, it may supply a cheaper universal action.
We do not split one maximum-delay batch into multiple services while
pretending that its delay charge stays unchanged.

Both models inherit $e/(e-1)$ from the same theorem, with no additional
temporal loss. Section~\ref{ag:shared:sec:applications} verifies the
actions and costs. The matching example is already a single positive
edge. The general theorem needs neither a hierarchy of requirements
nor submodularity; its two explicit spatial conditions are sufficient.

%% file: sections/facility_overview.tex
\section{Facilities: connections change the comparison}
\label{sec:facility-story}
A facility shares its opening among requests, but charges a separate
connection distance to each occurrence. If $k$ requests lie at distance
$d>0$ from a facility of opening cost $c$, serving them together costs
$c+kd$ before delay. No fixed spatial bound independent of $k$ is
available in this example. The universal-action theorem therefore
cannot simply be applied to facility location.

We instead compute an assignment from a finite observed initial
history and keep it fixed thereafter. The initial requests are served
immediately, and all subsequent decisions are deterministic functions
of the observed input and known instance data. The offline optimum
remains free to choose a different facility for every request.
The resulting deterministic long-run ratio is $3$, with arbitrary
nonnegative opening costs (Theorem~\ref{fl:max:thm:main}).
The expectation is over input arrivals alone.

\subsection{Separate connections from opening and delay}
Fix an assignment $\pi$ and let $\Lambda_f$ be the rate assigned to
facility $f$. Its expected connection cost rate is
$\sum_x\lambda_xd(x,\pi(x))$, regardless of service times.
For opening plus maximum delay, immediate service has rate
$c_f\Lambda_f$, whereas long collection periods approach rate one.
The assignment should therefore make the following quantity small:
$$\Psi(\pi)=\sum_x\lambda_xd(x,\pi(x))+
\sum_f\min\{c_f\Lambda_f,1\}.$$
This is a target cost rate, not yet a complete service policy.

A useful reformulation chooses which facilities will support shared
batches first. Each chosen facility contributes one unit to the target
rate. A client can either connect to one of those facilities or use
its cheapest individual service. This gives a static location problem
that includes both service modes. The equivalence is proved in
Lemma~\ref{fl:max:lem:surrogate}; it is not an assumption that all
facility opening costs equal one.

\subsection{Choose an assignment from observed arrival gaps}
Let $b_x$ be the distance from type $x$ to its nearest facility and let
$s_x=\min_f\{c_f+d(x,f)\}$ be its cheapest individual service cost.
List the positive-rate types in a fixed order $x_1,\ldots,x_m$ and put
$\Lambda=\sum_x\lambda_x$.
Serve the first $m$ requests of the merged arrival stream immediately,
recording their times $A_1,\ldots,A_m$ and setting $A_0=0$.
At time $\sigma=A_m$, compute
$$E_{x_j}=\frac{\Lambda}{\lambda_{x_j}}(A_j-A_{j-1}),\qquad 1\le j\le m.$$
The type index $x_j$ is fixed beforehand and need not be the type of
the $j$th actual request. The merged gaps are independent exponentials
of rate $\Lambda$, so the scaled values $E_x$ are independent
exponentials of their respective rates $\lambda_x$.
No random number is generated by the policy.

Set $r_x=b_x+E_x$, keep the candidate only when $r_x<s_x$, and process
the candidates in increasing order of $r_x+b_x$.

A candidate represents the facility set
$A(x,r_x)=\{f:d(x,f)<r_x\}$.
Accept it if this entire set is disjoint from all previously accepted
facility sets. Each accepted candidate selects a nearest facility to
its center. The test uses the full facility sets, so it can detect a
conflict at a facility that neither candidate selects.
Finally, each type connects to the nearest selected facility if this
is cheaper than its individual service, and uses its cheapest
individual service otherwise.

The construction uses only the first $m$ observed gaps.
If the input contains fewer than $m$ requests, all of them have already
been served immediately. Conditional on the initialization history,
the stream after $\sigma$ is a fresh Poisson input with the original
rates (Lemma~\ref{fl:max:lem:observed-seeds}).
Algorithm~\ref{fl:alg:maximum} gives the complete deterministic policy,
and Theorem~\ref{fl:max:thm:main} states its complexity.

\subsection{Compare the same packing with the offline optimum}
To analyze the assignment, consider an auxiliary Poisson process of
radii on $(0,s_x)$ for each type, with intensity $\lambda_x$.
Apply the same greedy rule, counting accepted empty facility sets as
well. Let $\mu$ be the expected number of accepted points and write
$B_0=\sum_x\lambda_xb_x$.
Empty sets do not affect the selected facilities. For any type, only
its first nonempty candidate can affect the selection, so the output
has exactly the marginal assignment law of the observed-gap construction above.

The expected number of selected facilities is $\mu-B_0$.
Poisson insertion bounds the expected connection and individual-service
charges by $B_0+2\mu$. Adding the two contributions gives
$$\E[\Psi(\pi)]\le3\mu.$$
The insertion argument uses the priority $r_x+b_x$: a rejected ball
shares a facility with an earlier accepted ball, whose selected
facility is close enough to bound its connection cost.

The same packing count also gives an offline lower certificate.
Request $i$ is active for $s_{x_i}$ time units after its arrival.
At time $\tau$ in this interval, give it residual radius
$s_{x_i}-(\tau-a_i)$ and apply the same greedy rule.
Within a batch served at facility $f$, at most one accepted ball
contains $f$ at any time. The opening and the batch's maximum wait
pay for the integrated contribution of these portions. Each request's
remaining portion has length at most its connection distance to $f$.
Consequently, the integrated accepted count lower-bounds every offline
schedule, even if that schedule changes facilities between occurrences.

Stationarity identifies the expected interior time-slice count with
$\mu$. Concatenating optimal schedules for consecutive equal-length
blocks removes the initial boundary loss and gives
$$\E[\OPT_T]\ge T\mu\qquad\text{for every }T>0.$$
Together, these comparisons imply
$$T\E[\Psi(\pi)]\le3\E[\OPT_T].$$
Section~\ref{sec:facility} proves both comparisons in full.
The assignment inequality averages over the observed initial gaps.
It uses only their marginal law and does not require independence
between the assignment and the full input.

\subsection{Realize the target rate and finish the input}
After initialization, facilities with $c_f\Lambda_f\le1$ serve their
assigned requests immediately. The others use predetermined service
times measured from $\sigma$, clearing their entire queues and skipping
empty batches. The calendar has successive gap lengths
$c_f,2c_f,3c_f,\ldots$. Over a long horizon, the number of openings is
sublinear, while the sum of maximum-wait charges is bounded by the
elapsed calendar time. This realizes the target unit rate for each
such facility.

Initialization has expected cost at most $mL/\Lambda$, where
$L=\sum_x\lambda_xs_x$.
The remaining horizon $h=(T-\sigma)_+$ and the assignment can be
correlated, but nonnegativity gives
$$\E[h\Psi(\pi)]\le T\E[\Psi(\pi)]\le3\E[\OPT_T].$$
This is the step that transfers the assignment comparison to the
policy using observed arrivals. The proof never factors this expectation
into a product of expectations.

Every request has a finite next service time. The final clearing
contributes a correction of order $O_I(\sqrt T+1)$, which is included
before taking the long-run ratio. The exact factor $3$ therefore
does not give a fixed bound on all individual waits.
Lemma~\ref{fl:max:lem:scheduled} proves the complete accounting.
If bounded waiting is preferred, fixed periods $c_f/\varepsilon$ at
the scheduled facilities give factor $3(1+\varepsilon)$ for any
$\varepsilon>0$, with explicit constant initialization and completion terms
(Corollary~\ref{fl:max:cor:bounded-wait}).

\paragraph{Guide to the technical development.}
The shared-service argument needs an interval lower bound, a calendar
upper bound, and the one-location sharpness proof. The facility result
adds an observed-gap initialization, an assignment bound, a time-slice
packing certificate, and an explicit calendar realization. The following sections give
these arguments in full, beginning with their probability tools.
The shared factor is optimal over its class; no matching optimality
claim is made for the facility factor.

%% file: sections/technical_tools.tex
\section{Probability tools for the proofs}\label{sec:technical-tools}
The preceding sections described the algorithms and the comparisons
behind their guarantees. We now give the complete technical arguments.
This section collects the Poisson identities used to average the
lower-bound constructions; subsequent sections prove the model-specific
estimates. The insertion identity counts the contribution of a point
whose acceptance depends on the other sampled points. In the facility
proof it is applied to an auxiliary Poisson process of radii, sampled
independently of the actual arrivals. It is an analysis tool, not
information supplied about future requests.
The facility proof also uses the independent exponential gaps of a
merged stream to compute its assignment from an observed initial history.

We write $z_+=\max\{z,0\}$, use $\mathbf1_A$ for the indicator of an
event, and let $|J|$ denote the length of a measurable time set.
Charging intervals are half-open. Nonnegative terms permit exchanges
of sums, integrals, and expectations by Tonelli's theorem.

\subsection{Poisson identities used in the proofs}\label{sec:poisson}
For a rate-$\lambda$ process, the number $N(J)$ of arrivals in an
interval $J$ has distribution
\[
 \Pr(N(J)=k)=e^{-\lambda|J|}\frac{(\lambda|J|)^k}{k!},
 \qquad \E[N(J)]=\operatorname{Var}(N(J))=\lambda|J|.
\]
Counts in disjoint intervals are independent. Conditional on $n$
arrivals in a finite interval, their unordered times are $n$
independent uniform points in that interval. In particular, for a
nonnegative deterministic function $g$,
\begin{equation}\label{eq:poisson-sum}
 \E\left[\sum_{a_i\in[0,T]}g(a_i)\right]
   =\lambda\int_0^T g(t)\,dt.
\end{equation}
Conditioning on the count proves this identity: the conditional
expectation is $nT^{-1}\int_0^Tg(t)\,dt$, and $\E[n]=\lambda T$.
We use it, for example, with $g(t)$ equal to the wait until the next
deterministic clock tick. Nonnegative integrands also allow us to
exchange sums, integrals, and expectations by Tonelli's theorem.

For independent streams, superposition adds their rates. Conversely,
independently assigning an arrival a mark $j$ with probability $q_j$
produces independent Poisson streams of rates $q_j\lambda$. The
generating function
$\E[\prod_j z_j^{N_j}]=\exp(\lambda|J|\sum_jq_j(z_j-1))$
factors into the generating functions of these independent counts.
These statements apply to marks chosen independently for each
occurrence; arbitrary routing based on the future need not preserve
a Poisson law.

The void probability is $\Pr(N(J)=0)=e^{-\lambda|J|}$.
It remains valid for a deterministic interval that depends on the
instance parameters. We will not apply it to intervals selected by
an offline schedule after observing arrivals. Instead, our lower
bounds first establish an inequality for every realized schedule
and then average a certificate whose counting regions are fixed
independently of that schedule.

We also use the following extension of~\eqref{eq:poisson-sum}, in
which the contribution of a distinguished arrival depends on the
other arrivals. The same identity on intervals of radii will be used
for the facility packing construction.

\begin{lemma}[Poisson insertion]\label{fl:lem:insertion}
Fix $x$, and let $u_x(a,J)$ be a nonnegative measurable quantity defined
for a distinguished type-$x$ request at time $a$ and a finite multiset $J$
of other requests. Then
\begin{equation}\label{fl:eq:insertion}
 \E\!\left[\sum_{i\in I_T:\,x_i=x}
   u_x(a_i,I_T\setminus\{i\})\right]
 =\lambda_x\int_0^T\,\E[u_x(a,I_T)]\,da.
\end{equation}
On the right, $I_T$ is an independent Poisson input to which the
distinguished request can be added inside the definition of $u_x$.
\end{lemma}
\begin{proof}
Condition on the number $N_x=n$ of type-$x$ arrivals. Their times are
independent uniforms in $[0,T]$, and exchangeability makes the sum on
the left equal in expectation to $n$ times the contribution of one
distinguished uniform time, with $n-1$ other type-$x$ times. The identity
$n\Pr(N_x=n)=\lambda_xT\Pr(N_x=n-1)$ changes the latter $n-1$ points
back into an unconditional type-$x$ Poisson process. All other types
retain their independent laws. Integrating the distinguished uniform
time proves~\eqref{fl:eq:insertion}. No conditioning on an arrival at one
prescribed time is needed.
\end{proof}

\paragraph{Arrivals observed by a causal policy.}
In the sharpness proof we count arrivals finding the system empty.
The relevant identity is the predictable form of
\eqref{eq:poisson-sum}: for a bounded nonnegative process $H(t)$
determined just before time $t$,
\[
 \E\left[\sum_{a_i\in[0,T]} H(a_i)\right]
   =\lambda\,\E\left[\int_0^T H(t)\,dt\right].
\]
For a simple predictable process, constant on deterministic intervals
with coefficients determined at their left endpoints, this follows
from independent increments. Approximation by such processes gives
the identity for predictable integrands. In particular it applies
to the left-limit indicator of an empty queue under a causal policy,
including a policy with independent internal randomness. This does
not assert the identity for a state chosen using future arrivals.

%% file: modules/ag/sections/maximum.tex
\section{A sharp theorem for shared services}\label{ag:sec:maximum}
We develop the core result in three steps. First, one location
reveals an exact description of offline cost and the two service
regimes. Next, singleton costs extend the lower bound to general
shared services, and a universal action supplies the upper bound.
Finally, we show that the resulting constant cannot be improved
even by randomization. The same development keeps track of
completion costs, which matter when a busy instance uses a long
service period.

\subsection{One location: the cost of combining requests}\label{sec:warmup}
Consider one request location with rate $\lambda>0$ and a service
cost $c>0$, independent of the number of pending requests. This is
a one-edge aggregation problem, or a single facility colocated
with its clients. There is no spatial choice, but the one-location
instance already exhibits the interval certificate and the timing
tradeoff used in the general theorem.

Now the delay of a nonempty batch is the waiting time of its oldest
request. Immediate service costs $c\lambda T$ in expectation. A
fixed-period policy with period $h$ has at most $T/h+1$ openings, and each nonempty
batch has delay at most $h$. Hence its total cost is at most
$(T/h+1)(c+h)$. By taking $h$ large, its long-run upper rate approaches
$1$, independently of $c$ and $\lambda$. This motivates the proxy
\begin{equation}\label{fl:eq:warm-max-proxy}
 m_c(\lambda)=\min\{c\lambda,1\}.
\end{equation}
The upper rate $1+c/h$ approaches one as the period grows. To see
the exact behavior, skip empty periods. A complete period has expected
opening cost $c(1-e^{-\lambda h})$ and expected maximum waiting
\[
 \int_0^h(h-u)\lambda e^{-\lambda u}\,du
 =h-\frac{1-e^{-\lambda h}}{\lambda}.
\]
Thus its exact long-run cost rate is
\[
 R(h)=1+\left(c-\frac1\lambda\right)
                  \frac{1-e^{-\lambda h}}h.
\]
In the busy regime $\lambda c>1$, every finite period has rate strictly
above one; at $\lambda c=1$, every finite period has rate exactly one.
For $\lambda c<1$, it is below one. In all three cases, completing a
finite input may require a service after $T$, whose cost must still
be included in the finite-horizon bound.

Here is the intuitive reason for the difference. Under additive delay,
each extra arrival contributes its own waiting time. Making the period
longer therefore becomes increasingly expensive when the stream is busy.
Under maximum delay, a hundred later arrivals may join the oldest one
without increasing the delay charge of that batch. The policy may
therefore use a much longer period. The resulting rate calculation is
simple, but it says little yet about an offline schedule that chooses
batch boundaries after seeing all arrivals.

\paragraph{The one-site offline optimum.}
For maximum delay, the one-site offline optimum has a particularly useful
samplewise description. If arrival times are $a_1<\cdots<a_m$, then
\begin{equation}\label{fl:eq:warm-max-exact}
 \OPT(I_T)
 =\left|\bigcup_{i=1}^m[a_i,a_i+c]\right|.
\end{equation}
Each arrival draws an interval of length $c$ to its right. The union
records precisely how much of the time line can be covered by grouping
nearby arrivals: a run of overlapping intervals is served as one batch,
while a gap larger than $c$ starts a new batch. This picture explains
the identity before any expectation is taken.
To prove the lower inequality, a batch served at $t$ with earliest
arrival $a_-$ has all intervals of its requests inside $[a_-,t+c]$;
the latter interval has length $c+t-a_-$, exactly the batch cost.
Union measure is subadditive across batches. For the reverse inequality,
sort the arrivals and place consecutive ones in the same batch whenever
their gap is at most $c$. Serve each batch at its last arrival time.
Its cost is $c$ plus the span from its first to its last arrival, which
equals the length of that connected component of the interval union.
Summing gives~\eqref{fl:eq:warm-max-exact}.

For times at least $c$ from the endpoints of $[0,T]$, the probability
that a fixed point is covered by this union is $1-e^{-\lambda c}$.
The two boundary pieces have total length at most $2c$, so integrating
the coverage indicator and taking expectations gives
\begin{equation}\label{fl:eq:warm-max-offline-rate}
 \lim_{T\to\infty}\frac{\E[\OPT(I_T)]}T
 =1-e^{-\lambda c}.
\end{equation}
This explains why $\min\{c\lambda,1\}$ is a constant-scale
online proxy: for $z=\lambda c$,
$1-e^{-z}\ge(1-e^{-1})\min\{z,1\}$.
The inequality follows from the monotonic decrease of
$(1-e^{-z})/z$ on $(0,1]$ and monotonic increase of
$1-e^{-z}$ on $[1,\infty)$. We next extend the interval argument to
shared services using singleton spatial costs. Facility location will
require a further step, because different requests in one batch pay
separate connection distances.

\paragraph{A concrete interval example.}
Suppose $c=1$ and arrivals occur at $0$, $0.4$, and $2$.
The union is $[0,1.4]\cup[2,3]$, of length $2.4$.
Serving the first two requests at $0.4$ costs $1+0.4$, and
serving the last at $2$ costs $1$. Serving all three at $2$
would instead cost $1+2=3$. The gap between the second and third
arrivals makes another opening cheaper than continuing the first
batch. This example also separates the interval drawn for a
request from its actual waiting interval: the former has prescribed
length $c$, regardless of when the request is served.

\input{figures/one_site_example}

\paragraph{Which part generalizes?}
With several types, a batch may contain requests with different
singleton costs. There need not be a schedule whose cost equals
the union measure. The lower inequality, however, still holds:
each singleton cost is at most the spatial cost of its batch,
so all request intervals fit in one paid envelope. For the upper
bound we need a different ingredient, namely an action that can
serve all possible types for a fixed spatial cost. We state these
two requirements separately rather than assuming the one-location
identity remains exact.

\subsection{The shared-service model and its lower bound}
We formulate the argument for the following class of services, then
apply it to our two tree problems. Assume a stateless batch-service model
with nonnegative spatial costs and the following properties.
For each request type $x$, the minimum singleton spatial cost $s_x\ge0$ is finite
and attained. Every batch with spatial cost $C$ satisfies $C\ge s_x$ for
each request location $x$ in the batch. Finally, a fixed action can serve
any multiset of requests at all locations for spatial cost at most a
finite constant $M$. In particular, this cost does not grow with request
multiplicity. We first prove the statement in this abstract form and
then specify the two tree actions to which it applies.
Discard zero-rate types first. If none remain, use the empty schedule:
both costs are zero and no ratio is asserted. Below we assume at least
one positive-rate type remains.

\paragraph{Notation.}
Put
\[
 L=\sum_x\lambda_xs_x,\qquad s_* =\max_x s_x,\qquad
 p=1-e^{-L},\qquad \rho_* =\frac{1}{1-e^{-1}}.
\]
The number $L$ is the expected singleton spatial cost arriving per
unit time. It is not itself a lower bound on offline cost, since
several requests may share one service. The interval certificate
below converts these singleton costs into a lower bound without
charging any spatial purchase twice. Its construction is independent
of how an offline schedule later chooses to group the requests.

\begin{lemma}[Interval-union certificate]
\label{ag:shared:lem:interval}
For every finite input and every feasible schedule $S$,
\[
 \left|\bigcup_i[a_i,a_i+s_{x_i}]\right|
 \le \cost(S).
\]
Consequently,
\begin{equation}\label{ag:shared:eq:shared-lb}
 \E[\OPT(I_T)]\ge
 \int_0^T\left[1-\exp\left\{-\sum_x\lambda_x\min(s_x,t)\right\}\right]dt
 \ge p(T-s_*)_+.
\end{equation}
\end{lemma}
\begin{proof}
A service at time $t$ with spatial cost $C$ and earliest arrival $a_-$
has every one of its request intervals inside $[a_-,t+C]$. That interval
has length $C+t-a_-$, the service cost. Subadditivity of union measure
proves the samplewise inequality, simultaneously for all schedules.

In more detail, each member $i$ arrived no earlier than $a_-$ and is
served at $t\ge a_i$. The assumption $s_{x_i}\le C$ gives

\[
 a_i+s_{x_i}\le t+C,
\]
 so its whole interval lies in that one
envelope. We take a union inside each batch and then across batches.
Overlapping envelopes only reduce the measure of the final union;
the presence of several roots cannot create an additional charge.

At a time $t\in[0,T]$, absence of coverage means that every type $x$ has
no arrival in $[\max(0,t-s_x),t]$. Independence and the Poisson void
probability give the displayed integrand. For $t\ge s_*$ it equals $p$.
The intervals intersected with $[0,T]$ have length $\min(s_x,t)$,
which explains the boundary term in the first integral rather than
silently replacing it by a stationary rate at time zero. Integrating
only over $[s_*,T]$ gives $p(T-s_*)_+$.
This argument never conditions on OPT's service times.
\end{proof}

Figure~\ref{ag:shared:fig:interval} illustrates the only samplewise step:
each request interval fits inside the envelope paid by the service that
contains it. Intervals from different roots or disconnected portions of
the tree are united only \emph{after} this containment is proved. Their
overlaps can only reduce the measure, so neither several roots nor a
shared edge is charged repeatedly.
\input{modules/ag/figures/max_interval}

\subsection{Deriving the online policy}
The lower bound has rate $p=1-e^{-L}$. What rate can a simple
policy attain? Immediate service costs exactly $L$ in expectation
per unit time. The ratio $L/(1-e^{-L})$ increases from $1$ to
$\rho_*$ as $L$ increases from $0$ to $1$. Thus no batching is
needed in this range. If $L>1$, immediate service can be expensive,
but the target rate $\rho_*p$ is strictly greater than one.

A calendar with period $H$ pays at most $M$ spatially and $H$
in maximum delay per nonempty tick. Its asymptotic upper rate is
therefore $1+M/H$. We choose the period by solving
\[
 1+M/H=\rho_*p,
 \qquad\text{giving}\qquad H=\frac{M}{\rho_*p-1}.
\]
The denominator is positive exactly in the busy regime $L>1$.
At $L=1$ it would vanish, but immediate service already attains
the desired factor there. This explains both the threshold and
the period, including why an infinite period is never needed.
Skipping an empty tick only improves the cost bound.

The policy depends on the instance before execution and on the
current queue during execution. It does not inspect the horizon
or predict a future arrival. An expensive universal action can
make $H$ large, which is why we will also state its finite-horizon
correction rather than interpreting the asymptotic ratio as a
uniform short-horizon guarantee.

\begin{algorithm}[!htbp]
\caption{Maximum-wait service for a finite collection of request types}
\label{ag:shared:alg:calendar}
\begin{algorithmic}[1]
\State Discard zero-rate types.
\If{no types remain}
  \State \Return the empty service schedule.
\EndIf
\State Compute each singleton spatial cost $s_x$, then
  $L=\sum_x\lambda_xs_x$ and $s_*=\max_xs_x$.
\State Serve zero-cost types immediately and remove them from the queue.
\If{$L\le1$}
  \State Serve each arriving positive-cost request immediately by its
    cheapest singleton action.
\Else
  \State Fix a universal action of cost at most $M$ and put
    $p=1-e^{-L}$, $H=M/(\rho_*p-1)$.
  \State Queue positive-cost arrivals until the next multiple of $H$.
  \Statex \textbf{At each positive multiple of $H$:}
    \If{the queue is nonempty}
      \State Use the universal action once on every queued request;
        clear the queue.
    \EndIf
\EndIf
\end{algorithmic}
\end{algorithm}

\subsection{The guarantee, including completion costs}
\begin{theorem}[Sharp shared-connection guarantee]\label{ag:shared:thm:shared}
For every fixed instance with $L>0$, there is a horizon-unaware
online algorithm with
\[
 \RoE(\ALG)\le\rho_* =\frac e{e-1}.
\]
For each fixed instance with $L>0$, the algorithm can use a finite, fixed
maximum waiting time. The universal long-run factor $\rho_*$ is optimal
over this model class, already for a single request location.
When $L=0$, every positive-rate type has free singleton service and
both costs are identically zero; no ratio is defined.
\end{theorem}
\begin{proof}[Upper bound and explicit finite-horizon correction]
In every regime, requests with $s_x=0$ are served immediately at zero
cost and excluded from the pending queue. If $L=0$, all positive-rate
requests have free singleton service. If
$0<L\le1$, serve each request immediately at singleton cost. Thus
$\E[\ALG(I_T)]=LT$. The function $L/(1-e^{-L})$ is increasing for $L>0$:
its derivative has numerator $1-(1+L)e^{-L}>0$. Hence
\[
 \E[\ALG(I_T)]\le \rho_*\,\E[\OPT(I_T)]+L s_*.
\]

If $L>1$, then $M>0$ and $\delta=\rho_*p-1>0$. Choose the fixed period
\[
 H=\frac{M}{\rho_*p-1}.
\]
At $H,2H,3H,\ldots$, use the universal action on all pending
positive-cost requests,
skipping empty batches. Every batch costs at most $M+H$. On input $I_T$,
all requests are served by $\lceil T/H\rceil H$, without an end signal.
Therefore, pathwise,
\[
 \ALG(I_T)\le (T/H+1)(M+H)=\rho_*pT+M+H.
\]
Combining this with \eqref{ag:shared:eq:shared-lb} gives
\begin{equation}\label{ag:shared:eq:shared-finite}
 \E[\ALG(I_T)]\le\rho_*\,\E[\OPT(I_T)]
       +M+H+\rho_*p s_*.
\end{equation}
The correction is independent of $T$ but depends on the fixed instance.
Dividing by the linearly growing optimum proves the long-run claim.
No period tending to infinity during execution is needed for this theorem.
\end{proof}

\subsection{Sharpness of the universal factor}

The upper bound compares two explicit policies with interval
coverage. To prove that the constant is inherent, we must compare
\emph{every} causal policy with the offline optimum. We return to
one location with $\lambda c=1$, the transition between the two
regimes. The offline rate is already known from
\eqref{fl:eq:warm-max-offline-rate}. The remaining task is to
show that every online algorithm pays at least one unit per unit
time in expectation.

The reason is a balance between two states. While its queue is
empty, arrivals occur at rate $\lambda$ and start new busy
periods. Each such period eventually requires a service costing
$c$. While its queue is nonempty, its oldest request accumulates
maximum waiting at unit rate. Since $\lambda c=1$, either state
contributes at the same expected rate. The proof below first
justifies clearing the entire queue and then makes this accounting
precise, including a busy period unfinished at $T$.

\begin{proof}[Sharpness of Theorem~\ref{ag:shared:thm:shared}]
Consider one location with spatial service cost $c>0$ and rate $\lambda$,
where $\lambda c=1$. For arrivals $a_1<\cdots<a_n$, the offline optimum is
\[
 0\quad(n=0),\qquad
 c+\sum_{i=2}^n\min(c,a_i-a_{i-1})\quad(n\ge1).
\]
Indeed, consecutive batches suffice: between consecutive requests one
chooses independently either to span the gap or to pay another opening.
Alternatively, this expression is precisely the measure of
$\bigcup_i[a_i,a_i+c]$, so the interval certificate and this batching
construction establish exact equality. The Poisson void calculation,
with at most $c$ of initial and terminal boundary length, yields
\[
 \lim_{T\to\infty}\frac{\E[\OPT(I_T)]}T=1-e^{-1}.
\]

It suffices to consider algorithms that clear all pending requests at
every opening. To justify this for the maximum-delay objective, retain
any original causal algorithm's opening times and move each request to
the first opening at or after its arrival. An original batch may split among
earlier openings, but its fragments' waiting intervals are disjoint
subintervals of its original waiting interval. Hence their summed
maximum delays do not exceed the original delay. Merging fragments at
an opening reduces this sum, and the number of openings does not grow.

The transformation is causal by simulating the original virtual state.
Empty openings can be deleted.

For a full-clear algorithm, let $Z(t)$ indicate a pending request, let
$A(T)$ count arrivals finding the system empty, and let $K(T)$ count
nonempty openings by $T$. Then $Z(T)=A(T)-K(T)$, starting empty.

Each period with $Z(t)=1$ runs from the first arrival after an
opening to the next opening. Its length is exactly that opening's
maximum waiting charge. The integral through $T$ also counts the
elapsed waiting of a final unfinished batch, whose eventual full
delay can only be larger.

The independent-increment Poisson compensator identity uses the
predictable empty-state indicator $1-Z(t-)$ and gives
\[
 \E[A(T)]=\lambda\,\E[\int_0^T(1-Z(t-))\,dt].
\]
The left and right versions of $Z$ differ only at jump times, which
form a measure-zero set for the time integral below.
Using $\lambda c=1$, the opening cost through $T$ plus holding accrued
through $T$ equals, in expectation,
\[
 c\,\E[K(T)]+\E[\int_0^TZ(t)\,dt]=T-c\,\E[Z(T)].
\]
If $Z(T)=1$, completion requires an additional opening costing $c$;
future waiting is nonnegative. Thus the full expected cost is at least
$T$ for every causal algorithm. This also holds for randomized algorithms
with independent randomness and even when $T$ is announced in advance.
If expected cost is infinite the lower bound is immediate. Dividing by
the offline rate proves the sharp factor $\rho_*$.
\end{proof}

\subsection{A limitation at finite horizons}\label{ag:shared:sec:finite}
\begin{proposition}[No uniform finite-horizon ratio for this algorithm]
The algorithm of Theorem~\ref{ag:shared:thm:shared}, including implementations that
reduce the spatial service to the actual pending locations, has unbounded
ratio on some fixed finite horizons as the instance varies.
\end{proposition}
\begin{proof}
Take a path rooted at $0$, with request locations $1$ and $M>1$,
rates $2$ and $M^{-3}$, and no activation charge. Then
$s_1=1,s_M=M$, $L=2+M^{-2}$, and universal spatial cost is $M$.
The algorithm's period is $H=M/(\rho_*(1-e^{-L})-1)$, with $H/M\to e$.
Take $T=1$ and sufficiently large $M$ so that $H>1$. The event that
there is exactly one near request and no far request has probability

\[
 2e^{-(2+M^{-3})}.
\]
 Conditional on this event its arrival is uniform in
$[0,1]$, and its cost is at least $1+H-a$, with conditional expectation
$H+1/2$. On the other hand, immediate singleton service gives
$\E[\OPT(I_1)]\le2+M^{-2}$. Thus
\[
 \frac{\E[\ALG(I_1)]}{\E[\OPT(I_1)]}
 \ge\frac{2e^{-(2+M^{-3})}(H+1/2)}{2+M^{-2}}
 \longrightarrow\infty.
\]
This disproves a uniform finite-horizon claim for this algorithm, not the
existence of some other algorithm with such a guarantee.
\end{proof}

The theorem is therefore a statement about each fixed instance as
$T$ grows. It does not assert an instance-uniform horizon at which
the boundary term becomes small. The spatial hypotheses also
remain essential: the abstract theorem does not supply algorithms
for computing cheapest singleton actions, and a persistent server
or a per-occurrence connection sum requires further analysis.

%% file: figures/one_site_example.tex
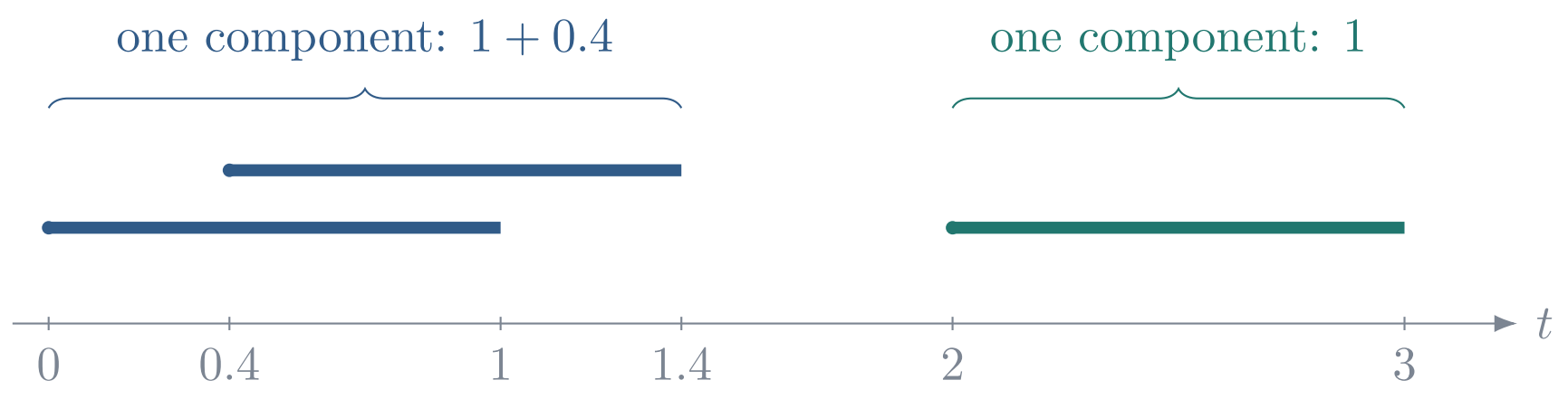
\begin{figure}[t]
\centering
\begin{tikzpicture}[x=3.3cm,y=0.7cm,>=Latex,font=\small]
\draw[->,PaperInk!60] (-.08,0)--(3.25,0) node[right] {$t$};
\foreach \x/\lab in {0/0,.4/0.4,1/1,1.4/1.4,2/2,3/3}
  {\draw[PaperInk!60] (\x,.07)--(\x,-.07) node[below] {$\lab$};}
\draw[line width=2.5pt,PaperBlue] (0,1)--(1,1);
\draw[line width=2.5pt,PaperBlue] (.4,1.6)--(1.4,1.6);
\draw[line width=2.5pt,PaperTeal] (2,1)--(3,1);
\fill[PaperBlue] (0,1) circle (1.4pt);
\fill[PaperBlue] (.4,1.6) circle (1.4pt);
\fill[PaperTeal] (2,1) circle (1.4pt);
\draw[decorate,decoration={brace,amplitude=4pt},PaperBlue]
 (0,2.25)--(1.4,2.25) node[midway,above=6pt] {one component: $1+0.4$};
\draw[decorate,decoration={brace,amplitude=4pt},PaperTeal]
 (2,2.25)--(3,2.25) node[midway,above=6pt] {one component: $1$};
\end{tikzpicture}
\caption{At one location with cost $c=1$, the three arrival intervals
have union measure $2.4$. The offline schedule serves each component
at its last arrival, here at $0.4$ and $2$. The intervals are
certificate intervals, not actual waiting intervals.}
\label{fig:one-site-example}
\end{figure}

%% file: modules/ag/figures/max_interval.tex
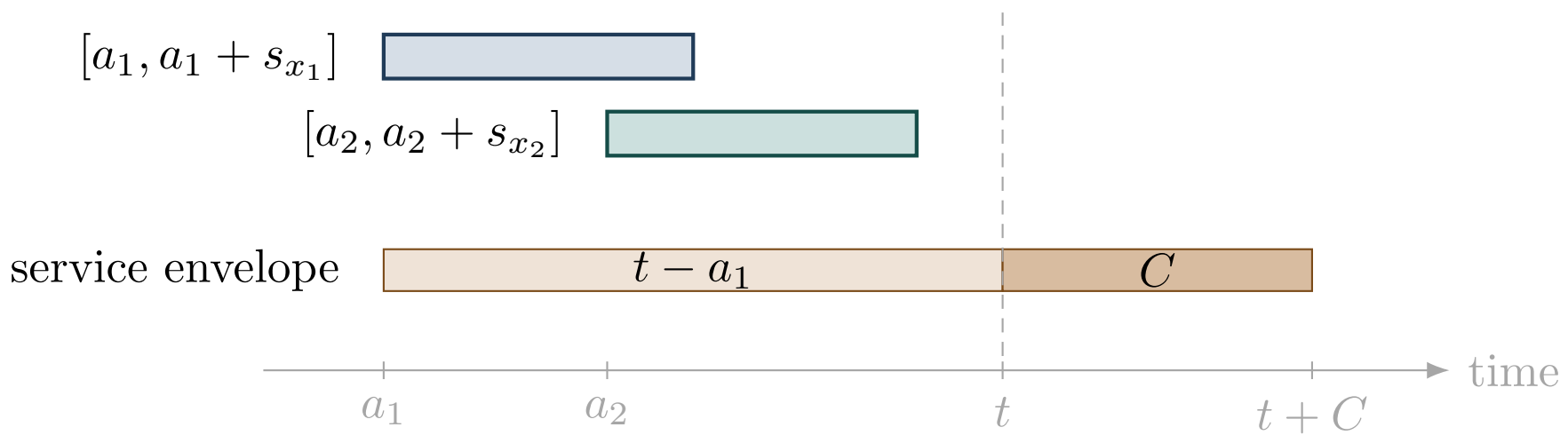
\begin{figure}[tbp]
\centering
\begin{tikzpicture}[x=1.28cm,y=.82cm,>=Latex,font=\small]
  \draw[->,gray!70] (0,0)--(6.9,0) node[right] {time};
  \foreach \x/\lab in {.7/a_1,2.0/a_2,4.3/t,6.1/{t+C}}
    {\draw[gray!70] (\x,.08)--(\x,-.08) node[below] {$\lab$};}
  \filldraw[fill=PaperBlue!20,draw=PaperBlue!65!black,thick]
      (.7,2.65) rectangle (2.5,3.05);
  \node[anchor=east] at (.55,2.85) {$[a_1,a_1+s_{x_1}]$};
  \filldraw[fill=PaperTeal!23,draw=PaperTeal!65!black,thick]
      (2.0,1.95) rectangle (3.8,2.35);
  \node[anchor=east] at (1.85,2.15) {$[a_2,a_2+s_{x_2}]$};
  \filldraw[fill=PaperAmber!18,draw=PaperAmber!75!black]
      (.7,.72) rectangle (4.3,1.10);
  \filldraw[fill=PaperAmber!43,draw=PaperAmber!75!black]
      (4.3,.72) rectangle (6.1,1.10);
  \node[anchor=east] at (.55,.91) {service envelope};
  \node at (2.5,.91) {$t-a_1$};
  \node at (5.2,.91) {$C$};
  \draw[densely dashed,gray!70] (4.3,0.13)--(4.3,3.25);
\end{tikzpicture}
\caption{The two singleton intervals overlap, but both lie inside
the envelope $[a_1,t+C]$ of their common service at time $t$.
The envelope has length $t-a_1+C$, exactly that service's cost.
Taking a union avoids charging the shared spatial cost twice.}
\label{ag:shared:fig:interval}
\end{figure}

%% file: sections/tree_applications.tex
\section{Verification of the tree applications}
\label{ag:shared:sec:applications}

We now allow the requests to specify spatial connections in a tree.
In multi-root aggregation the service may select a root for each
batch; in pair aggregation each request specifies two endpoints.
The common idea is to identify what one request costs alone and
what one fixed action costs when every possible type is present.
Once these two actions are available, Section~\ref{ag:sec:maximum}
already supplies the temporal algorithm and its full analysis.

\input{sections/tree_models}

If no positive-rate request type remains, use the empty schedule.
In the constructions below, we assume that at least one type remains.

\subsection{Multi-root aggregation}
\paragraph{Idea.}
For light demand, choose the cheapest root separately for each
arrival. For heavy demand, select a single root and a subtree
covering all possible clients, and share that action at calendar
ticks. The offline optimum retains its freedom to change roots;
the singleton interval certificate, rather than a restriction on
that optimum, pays for the online policy.

\paragraph{Singleton and universal actions.}
For multi-root aggregation, retain the original tree $\mathcal T$
and its terminal set $R$. A singleton at $x$ can be served from its
cheapest root, so
\[
 s_x=\min_{r\in R}\{c_r+d_{\mathcal T}(r,x)\}.
\]
Every service containing $x$ pays at least $s_x$: its chosen subtree
contains the entire root-to-$x$ path and pays that root's activation.
Let $X$ be the finite set of positive-rate client locations. Choose a
root $r_0$ minimizing
\[
 M_r=c_r+w\bigl(\operatorname{hull}_{\mathcal T}(\{r\}\cup X)\bigr),
 \qquad M=\min_{r\in R}M_r.
\]
At any nonempty tick, the subtree realizing $M$ can serve all queued
clients, regardless of multiplicity. This gives every hypothesis of
Theorem~\ref{ag:shared:thm:shared}. It also explains why the heavy-case
algorithm need not decide a different root for each batch: its universal
root is paid for by the interval-union lower bound in expectation.

\subsection{Pair aggregation}
\paragraph{Idea.}
A pair request must receive its entire endpoint path. The path
cost is therefore a lower bound on any batch serving it, even
when other paths overlap. A hull containing every possible
endpoint is a universal connected service. If disconnected
unions are legal as a single action, their smaller cost may be
used instead. We do not split one maximum-delay batch into
separate actions and assume that its delay remains unchanged.

\paragraph{Notation and actions.}
For pair aggregation, type $i$ requires the unique tree path
$P_i=P_{\mathcal T}(u_i,v_i)$. Its singleton spatial cost is
$s_i=w(P_i)$. Any legal service containing that type contains $P_i$,
so its spatial cost is at least $s_i$. A universal connected action
is the hull of all positive-rate endpoints,
\[
 M=w\bigl(\operatorname{hull}_{\mathcal T}
       (\{u_i,v_i:\lambda_i>0\})\bigr).
\]
It contains every required path. If the model permits a forest service,
one may instead use the union $\bigcup_iP_i$, with potentially smaller
$M$. Either action has cost independent of the number of queued
occurrences. Thus the same theorem applies without assigning a common
root to pair demands.

\subsection{Algorithm and guarantee}

Compute the corresponding singleton costs and universal action,
and run Algorithm~\ref{ag:shared:alg:calendar} without changing
its threshold or calendar. For light demand it uses the singleton
actions; otherwise it uses the chosen universal action on all
queued requests. Every batch dominates each member's singleton
cost, and the universal cost does not depend on multiplicity.
The shared-service theorem therefore gives the following result.
\begin{corollary}[Both maximum-delay tree extensions]
\label{ag:shared:cor:trees}
For each fixed finite tree instance with $L>0$,
Algorithm~\ref{ag:shared:alg:calendar}
has long-run RoE at most $e/(e-1)$ in multi-root aggregation and pair
aggregation with per-batch maximum waiting. Its finite-horizon correction
is $Ls_*$ in the light regime and $M+H+\rho_*ps_*$ in the heavy regime.
\end{corollary}

%% file: sections/tree_models.tex
\paragraph{Notation and service conventions.}
Let $\mathcal T=(V,E)$ be a finite tree with nonnegative edge costs.
In multi-root aggregation, requests occur at client vertices; a
service chooses one root $r$ from a finite nonempty set $R$ and a
connected subtree containing $r$ and every client in its batch.
It pays the subtree cost and a root activation $c_r\ge0$. Each
service purchases its network anew, and the offline optimum may
change roots between services.

In pair aggregation, a type has endpoints $u_i,v_i$ and requires
their unique path $P_i$. A connected service can buy the hull of
the paths of its batch. We also allow the variant in which one
service may purchase their union as a forest. The shared-service
theorem applies to either convention, with the corresponding
singleton and universal actions. A forest is not silently split
into several maximum-delay services: such a split may increase
the temporal charge.

%% file: sections/facility_proof.tex
\section{Proof of the facility guarantee}\label{sec:facility}
We now consider a batch served by opening a facility and connecting
each of its requests. The delay is still one maximum wait for the
batch, but each occurrence pays its own connection distance. This
changes the spatial part of the problem even when the set of
possible client locations is finite.

\paragraph{Idea and illustration.}
Suppose a client is at distance $d>0$ from the only facility.
Serving $k$ occurrences together costs $c+kd$ before delay.
There is consequently no universal action of bounded cost
independent of multiplicity. The shared-service theorem cannot
simply be applied with the facility as that action. Yet under a
fixed assignment, connection cost has a predictable rate,
$\lambda d$, and only the opening-plus-delay part benefits from
batching. This suggests first constructing a good assignment
and then using the one-facility service rule on each assigned
stream.

For the assignment, we scale observed arrival gaps from an initial
sequence of immediately served requests. These values have independent
exponential laws, and the future stream is independent of that observed
history. We use the resulting radii to greedily select balls whose
facility sets are disjoint. An auxiliary
Poisson sample makes the connection between this construction and
the offline lower bound explicit. At each time slice, apply the same
packing rule to shrinking balls associated with active requests.
Within any offline batch, the portions containing its facility are
paid by its opening and maximum wait, while the remaining portions
are paid by connections. Poisson insertion bounds the assignment
cost by three times the expected packing size. Stationarity and
offline subadditivity compare that size with unrestricted OPT.
We then account for initialization and give shifted service calendars
that include all completion costs. Every action of the policy is a
deterministic function of the observed input.

\input{modules/fl/sections/model}
\input{modules/fl/sections/maximum}

%% file: modules/fl/sections/model.tex
\subsection{The facility model}\label{fl:sec:preliminaries}

\paragraph{Notation.}
Let $(X\cup F,d)$ be a finite metric space, where $X$ contains possible
request locations and $F\ne\varnothing$ contains possible facility
locations. The sets may overlap. All distances are finite, and each
facility $f$ has a finite opening cost $c_f\geq0$. Nothing requires
opening costs to be uniform. Distinct requests at the same point remain
distinct occurrences. We write $\pos z=\max\{z,0\}$.

At time $t$, an algorithm may choose $f\in F$ and a nonempty subset $B$ of
pending requests whose arrival times $a_i$ satisfy $a_i\le t$. It pays
$c_f$ and connects each $i\in B$ to $f$, paying $d(x_i,f)$ for that
request. The facility closes after this event; opening it at another time
costs again. Several service events at one time are allowed, but each has
its own opening and delay charge. There is no capacity constraint or travel
time. \begin{equation}\label{fl:eq:max-cost}
\cost(B,f,t)=c_f+\sum_{i\in B}d(x_i,f)+\max_{i\in B}(t-a_i).
\end{equation}
The total cost is the sum over service events. The offline optimum
may choose different facilities for different occurrences of the same type.
The algorithm knows the metric, all opening costs, and all arrival rates.
Each event pays its own maximum-wait charge, even when several events
occur at the same time. The connection sum grows with request multiplicity;
there need not be a universal action of bounded spatial cost.

%% file: modules/fl/sections/maximum.tex
\subsection{A deterministic factor-three guarantee}\label{fl:sec:maximum}
We use the facility model of Section~\ref{fl:sec:preliminaries}.
The algorithm first serves a finite number of requests immediately.
It then computes an assignment from their observed arrival gaps and uses deterministic service times for each assigned stream.
An auxiliary Poisson packing bounds the assignment cost, and the same packing rule applied to time slices gives an offline lower certificate.

Discard zero-rate types and list the remaining types in a fixed order $x_1,\ldots,x_m$.
If $m=0$, use the empty policy.
For $m>0$, write
$$\Lambda=\sum_x\lambda_x,\qquad s_x=\min_f\{c_f+d(x,f)\},\qquad
b_x=\min_fd(x,f),\qquad L=\sum_x\lambda_xs_x.$$
For an assignment $\pi:X\to F$, define
$$\Lambda_f=\sum_{x:\pi(x)=f}\lambda_x,\qquad
\Psi(\pi)=\sum_x\lambda_xd(x,\pi(x))+\sum_f\min\{c_f\Lambda_f,1\},\qquad
\Phi=\min_\pi\Psi(\pi).$$
The opening-plus-delay term chooses between immediate service, of rate $c_f\Lambda_f$, and a calendar whose asymptotic rate is at most one.
The following theorem realizes this proxy after a finite initialization phase.
\begin{theorem}\label{fl:max:thm:main}
For every finite metric facility instance with nonnegative opening costs and independent Poisson arrivals, Algorithm~\ref{fl:alg:maximum} is deterministic.
When $m>0$, its completed cost satisfies, for every $T>0$,
$$\E[\ALG(I_T)]\le3\E[\OPT(I_T)]+\frac{mL}{\Lambda}
  +2\sum_{f\in F}\bigl(\sqrt{2c_fT}+c_f\bigr).$$
Consequently its long-run RoE is at most $3$ on every instance with positive optimal asymptotic cost.
The algorithm uses the first $m$ observed arrival gaps and $O(|X||F|+|X|\log|X|)$ arithmetic, comparison, and set operations to compute its assignment, with no auxiliary randomness.
\end{theorem}
The expectation is over the input arrivals alone.
The assignment becomes fixed after initialization; it is not chosen before arrivals begin.
The offline comparator remains unrestricted and may choose a different facility for each occurrence.
The operation count uses the real-arithmetic model with exact observed timestamps.

\subsection{The static proxy and the assignment algorithm}
Put $B_0=\sum_x\lambda_xb_x$.
Fix a nearest facility $n(x)$ and a singleton minimizer $g(x)$, so that
$d(x,n(x))=b_x$ and $c_{g(x)}+d(x,g(x))=s_x$.
All ties are broken by fixed orders.
For $S\subseteq F$, let
$$J(S)=|S|+\sum_x\lambda_x\min\{s_x,d(x,S)\},$$
with $d(x,\varnothing)=+\infty$.
The unit term for a selected facility represents the saturated opening-plus-delay rate; its physical opening cost remains $c_f$.
The following identity relates facility selection to fixed assignments.
\begin{lemma}\label{fl:max:lem:surrogate}
One has $\Phi=\min_{S\subseteq F}J(S)$.
For each $S$, assigning every type to a nearest facility in $S$ when $d(x,S)<s_x$, and to $g(x)$ otherwise, gives $\Psi(\pi)\le J(S)$.
\end{lemma}
\begin{proof}
For a fixed assignment, put in $S$ the facilities whose opening proxy uses the saturated term $1$.
Their assigned clients can connect within $S$ at no greater distance.
Every other client can use singleton fallback for cost at most its assigned opening-plus-connection charge.
Thus $J(S)\le\Psi(\pi)$.

Conversely, consider the assignment induced by $S$ in the statement.
Charge at most one for each facility in $S$ and charge $c_{g(x)}$ per fallback occurrence in expectation.
For each physical facility the actual opening proxy is no larger than the sum of these charges, including when it receives both direct and fallback assignments.
Adding connections gives $\Psi(\pi)\le J(S)$, proving the identity.
\end{proof}

The assignment map $G$ takes a positive vector $(e_x)$ as input.
For each type, form $r_x=b_x+e_x$ and retain it only if $r_x<s_x$.
Process the retained candidates in increasing order of $r_x+b_x$.
A candidate represents the facility set
$$A(x,r)=\{f\in F:d(x,f)<r\}.$$
Accept it when this entire set is disjoint from all previously accepted facility sets, and select its nearest facility $n(x)$.
Assign each type to its nearest selected facility when that distance is below $s_x$, and to $g(x)$ otherwise.
Disjointness is tested on the entire sets $A(x,r_x)$, not only on the selected nearest facilities.

The online policy supplies the vector using observed arrivals.
Let $A_j$ be the time of the $j$th request in the merged arrival stream, with $A_0=0$.
Serve the first $m$ requests immediately at their singleton minimizers.
At time $\sigma=A_m$, compute
$$E_{x_j}=\frac{\Lambda}{\lambda_{x_j}}(A_j-A_{j-1}),\qquad 1\le j\le m,$$
and set $\pi=G(E)$.
The index $x_j$ is the $j$th type in the fixed order, not necessarily the type of the $j$th arriving request.
This computation uses no internally generated random numbers.
If fewer than $m$ requests ever arrive, they have all been served immediately.

\begin{algorithm}[!ht]
\caption{Deterministic factor-three facility location with maximum delay}
\label{fl:alg:maximum}
\begin{algorithmic}[1]
\State Assign zero-rate types arbitrarily and discard them.
\If{no types remain}
  \State \Return the empty service schedule.
\EndIf
\State Fix the type order $x_1,\ldots,x_m$ and compute $\Lambda,s_x,b_x,n(x),g(x)$.
\State Set $A_0\gets0$.
\For{$j=1,\ldots,m$}
  \State On the next arrival of type $y$ at time $A_j$, serve it immediately at $g(y)$.
  \State Set $E_{x_j}\gets(\Lambda/\lambda_{x_j})(A_j-A_{j-1})$.
\EndFor
\State Set $\sigma\gets A_m$ and $r_x\gets b_x+E_x$ for each type.
\State Sort the candidates with $r_x<s_x$ by increasing $r_x+b_x$.
\State Set $U\gets\varnothing$ and $S\gets\varnothing$.
\For{each retained candidate $x$, in that order}
  \State $A_x\gets\{f\in F:d(x,f)<r_x\}$.
  \If{$A_x\cap U=\varnothing$}
    \State $U\gets U\cup A_x$ and $S\gets S\cup\{n(x)\}$.
  \EndIf
\EndFor
\State Assign $x$ to a nearest facility in $S$ if $d(x,S)<s_x$; otherwise set $\pi(x)=g(x)$.
\State Compute the assigned rates $\Lambda_f$.
\Statex \textbf{For each subsequent request at $x$:}
\If{$c_{\pi(x)}\Lambda_{\pi(x)}\le1$}
  \State Serve it immediately at $\pi(x)$.
\Else
  \State Enqueue it for $\pi(x)$.
\EndIf
\Statex \textbf{For $c_f\Lambda_f>1$, at times $\sigma+c_fj(j+1)/2$, $j=1,2,\ldots$:}
\State If the queue at $f$ is nonempty, open $f$ and serve the entire queue.
\end{algorithmic}
\end{algorithm}

\subsection{The law of the observed gaps}
For analysis, couple all finite inputs to an infinite Poisson stream.
This defines $\sigma$ and $\pi=G(E)$ even when $I_T$ has fewer than $m$ requests.
The actual algorithm does not access this continuation.
Let $\mathcal H$ be the history of the first $m$ merged arrivals, including their times and types.
The next lemma identifies both the marginal assignment law and the conditional law of the remaining input.
\begin{lemma}\label{fl:max:lem:observed-seeds}
The variables $E_x$ are independent and have distributions $\operatorname{Exp}(\lambda_x)$.
Conditional on $\mathcal H$, the request streams strictly after $\sigma$, translated by $-\sigma$, are independent Poisson streams of rates $(\lambda_x)$, with a law independent of $\mathcal H$.
\end{lemma}
\begin{proof}
Write $Z_j=A_j-A_{j-1}$ and let $M_j$ be the type of the $j$th merged request.
For independent exponential clocks, the next gap $Z$ and its winning type $M$ have joint density
$$\Pr(Z\in dz,M=x)=\lambda_xe^{-\Lambda z}\,dz
=\Lambda e^{-\Lambda z}\,dz\cdot\frac{\lambda_x}{\Lambda}.$$
Memorylessness restarts the independent clocks after each arrival.
Thus the $Z_j$ are independent $\operatorname{Exp}(\Lambda)$ variables, the $M_j$ are independent marks with probabilities $\lambda_x/\Lambda$, and the two sequences are independent.
For $u\ge0$,
$$\Pr(E_{x_j}>u)=\Pr\left(Z_j>\frac{\lambda_{x_j}}\Lambda u\right)
=e^{-\lambda_{x_j}u}.$$
The gaps and marks after index $m$ are independent of the first $m$ pairs and have their original laws.
This proves the conditional future statement, including independence across types by the same exponential-clock description.
\end{proof}

\subsection{An auxiliary Poisson packing}
For the analysis, independently sample a Poisson process of radii with intensity $\lambda_x\,dr$ on $(0,s_x)$ for each positive-rate type $x$.
This finite auxiliary sample is independent of the actual arrival processes.
Process all its points $(x,r)$ in increasing order of $r+b_x$, accepting a point precisely when $A(x,r)$ is disjoint from all earlier accepted facility sets.
Empty sets are always accepted and select no facility.
Every nonempty accepted set selects $n(x)$, which belongs to that set.
Let $S$ be the selected facility set, $K$ the number of accepted points including empty sets, and $\mu=\E[K]$.

This conceptual construction has the same marginal assignment law as $G(E)$ in Algorithm~\ref{fl:alg:maximum}, with $E$ defined on the infinite Poisson continuation.
Indeed, the nonempty sets of any one type are nested as the radius increases.
If its first nonempty point is accepted, it blocks every later point of that type.
If that first point is rejected, its earlier accepted blocker also intersects every larger set of that type.
Thus all later points are rejected and cannot affect the greedy output.
Empty points block nothing and may also be omitted.
The first point above $b_x$ is distributed as $b_x+\operatorname{Exp}(\lambda_x)$ and is present in the auxiliary interval exactly when it is smaller than $s_x$.
Lemma~\ref{fl:max:lem:observed-seeds} identifies this law with the vector computed from observed arrival gaps, proving the claimed equivalence.
Computing the minima, testing set intersections by scanning $F$, sorting at most $|X|$ candidates, and forming the assignment takes $O(|X||F|+|X|\log|X|)$ operations.

The nonempty accepted sets are pairwise disjoint and contain their selected facilities, so those facilities are distinct.
All sampled points with $r\le b_x$ have empty sets and are accepted.
Their expected number is $B_0$.
Consequently
$$\E|S|=\mu-B_0,\qquad \mu\le B_0+\min\{|X|,|F|\}.$$
The next lemma bounds connections through insertion into this auxiliary sample.
\begin{lemma}\label{fl:max:lem:assignment}
The assignment $\pi=G(E)$ satisfies $\E[\Psi(\pi)]\le3\mu$.
\end{lemma}
\begin{proof}
Let $p_x(r)$ be the probability that a distinguished point $(x,r)$ inserted into an independent auxiliary sample is accepted.
If it is rejected, an earlier accepted point $(y,u)$ shares a facility $h$ with it and has $u+b_y\le r+b_x$.
Insertion leaves every earlier greedy decision unchanged.
The nearest facility $n(y)$ chosen for the blocker therefore belongs to the original sample's set $S$.
The triangle inequality gives
$$d(x,n(y))\le d(x,h)+d(h,y)+b_y<r+u+b_y\le2r+b_x.$$
Hence
$$\Pr\{d(x,S)>2r+b_x\}\le p_x(r).$$
For an empty inserted set the inequality also holds, since it is always accepted.

Since $b_x\le\min\{s_x,d(x,S)\}$, integration of the survival function gives
$$\begin{aligned}
\E\sum_x\lambda_x\min\{s_x,d(x,S)\}
&=B_0+\sum_x\lambda_x\int_{b_x}^{s_x}\Pr\{d(x,S)>v\}\,dv\\
&\le B_0+2\sum_x\lambda_x\int_0^{(s_x-b_x)/2}p_x(r)\,dr\\
&\le B_0+2\sum_x\lambda_x\int_0^{s_x}p_x(r)\,dr
=B_0+2\mu.
\end{aligned}$$
The last equality is the Poisson insertion identity from Lemma~\ref{fl:lem:insertion}, applied to radii instead of arrival times.
Terms with $s_x=0$ vanish.
For $s_x>0$, conditioning on the Poisson count $N_x$ in $(0,s_x)$ gives independent uniform radii, and the identity
$j\Pr(N_x=j)=\lambda_xs_x\Pr(N_x=j-1)$ proves the same insertion formula on that interval.
Finally, $\E|S|=\mu-B_0$ and Lemma~\ref{fl:max:lem:surrogate} yield
$$\E[\Psi(\pi)]\le\E[J(S)]\le(\mu-B_0)+(B_0+2\mu)=3\mu.$$
\end{proof}

\subsection{The offline packing certificate}
We now apply the same greedy rule to a time slice of the actual requests.
The resulting certificate is valid for every realized schedule, independently of its facility choices.
\begin{lemma}\label{fl:max:lem:offline}
For every $T>0$, $\E[\OPT(I_T)]\ge T\mu$.
\end{lemma}
\begin{proof}
For a finite realization $I_T$ and time $\tau$, request $i$ is active when
$a_i\le\tau<a_i+s_{x_i}$ and then has residual radius
$$r_i(\tau)=s_{x_i}-(\tau-a_i).$$
Apply the auxiliary greedy rule to these active balls and let $K_T(\tau)$ count the accepted points, including empty sets.
At every time, at most one accepted set contains any given facility.
The integral of this count is a lower bound for every schedule $\mathcal S$:
$$\int_{\mathbb R}K_T(\tau)\,d\tau\le\cost(\mathcal S).$$
To prove this, fix a batch served at $f$ at time $t$, and let $a_-$ be its earliest arrival.
Split the active interval of request $i$ into a prefix of length $(s_{x_i}-d(x_i,f))_+$ and the remaining tail, whose length is at most $d(x_i,f)$.
During the prefix its ball contains $f$, apart from endpoints of measure zero.
The accepted prefix indicators of this batch sum to at most one at every time.
Every such prefix lies in $[a_-,t+c_f]$, because $s_x\le c_f+d(x,f)$.
Their integrated contribution is at most $c_f+t-a_-$, and the tails contribute at most the sum of connection distances.
Summing over batches proves the inequality, including batches completed after $T$.
Figure~\ref{fl:fig:maximum-intervals} illustrates the facility-set and time-slice constraints.

Let $s_* =\max_xs_x$ over positive-rate types, with $s_*=0$ if none remain.
For $\tau\in[s_*,T]$, each relevant arrival interval lies inside $[0,T]$.
Mapping a type-$x$ arrival to its residual radius gives a Poisson process of intensity $\lambda_x\,dr$ on $(0,s_x)$, independently across types.
Thus $\E K_T(\tau)=\mu$ throughout this interior interval, and Tonelli's theorem gives
$$\E[\OPT(I_T)]\ge(T-s_*)_+\mu.$$
No independence between different time slices is required.
The certificate is measurable: for finite $I_T$, its active sets and facility incidences change only finitely many times, and the relative priority of two simultaneously active points is constant.

Subadditivity removes the initial boundary term.
Put $V(T)=\E[\OPT(I_T)]$, which is finite since immediate singleton service has expected cost $T\sum_x\lambda_xs_x$.
Each offline batch can be served at its latest arrival without increasing cost, so an optimal schedule for a block can finish within that block.
Concatenating optimal schedules for $\ell$ consecutive blocks of length $T$ gives $V(\ell T)\le\ell V(T)$.
The blocks have identical Poisson laws, and deterministic endpoints contain no arrival almost surely.
Applying the interior lower bound at horizon $\ell T$ therefore yields
$$V(T)\ge\frac{(\ell T-s_*)_+}{\ell}\mu.$$
Letting $\ell\to\infty$ proves $V(T)\ge T\mu$ for every fixed $T>0$.
\end{proof}
\input{modules/fl/figures/maximum_intervals}

Together, Lemmas~\ref{fl:max:lem:assignment} and~\ref{fl:max:lem:offline} give
$$\Phi\le\E[\Psi(\pi)]\le3\mu\le3\frac{\E[\OPT(I_T)]}{T}.$$
This comparison uses only the marginal law of $\pi$.
It asserts neither a bound for every realized assignment nor independence of $\pi$ from the full input.
The next step conditions on the initialization history to use the independent future supplied by Lemma~\ref{fl:max:lem:observed-seeds}.

\subsection{Service calendars, initialization, and completion}
Let $C_{\mathrm{post}}(I_T)$ denote the cost after initialization, including completion after $T$, and set $h=(T-\sigma)_+$.
Conditional on $\mathcal H$, the values of $h$ and $\pi$ are fixed.
When $\sigma>T$, this cost is zero.
Otherwise the future stream at $f$ is Poisson of rate $\Lambda_f$.
Immediate service realizes the proxy when $c_f\Lambda_f\le1$.
For the remaining facilities, the increasing service times make the average opening rate vanish while the total maximum-delay rate is at most one.
The following estimate accounts for the final clearing.
\begin{lemma}\label{fl:max:lem:scheduled}
For every $T>0$, Algorithm~\ref{fl:alg:maximum} satisfies
$$\E[C_{\mathrm{post}}(I_T)\mid\mathcal H]\le h\Psi(\pi)
  +2\sum_{f:c_f\Lambda_f>1}\bigl(\sqrt{2c_fh}+c_f\bigr).$$
Every finite input is cleared without an end-of-input signal.
\end{lemma}
\begin{proof}
If $h=0$, the cost is zero and the bound holds.
For $h>0$, condition on $\mathcal H$ and measure time from $\sigma$.
The expected connection cost is $h\sum_x\lambda_xd(x,\pi(x))$.
An instant facility has opening-plus-delay rate $c_f\Lambda_f$.
For a scheduled facility, $c_f>0$.
Let $J_f$ be the first index with $t_{f,J_f}=c_fJ_f(J_f+1)/2\ge h$, and put $t_{f,0}=0$.
There are at most $J_f$ nonempty openings through completion.
Each batch at relative time $t_{f,j}$ waits at most $t_{f,j}-t_{f,j-1}=c_fj$, so the sum of maximum delays is at most $t_{f,J_f}$.
Moreover, $t_{f,J_f}-h\le c_fJ_f$ and $J_f\le\sqrt{2h/c_f}+1$.
Thus the opening-plus-delay cost at this facility is bounded pathwise by
$$c_fJ_f+t_{f,J_f}\le h+2c_fJ_f\le h+2\sqrt{2c_fh}+2c_f.$$
Adding all facilities proves the inequality.
Every request is served immediately or at its finite next calendar tick.
If initialization never finishes on a finite input, all its requests were already served immediately.
\end{proof}

\begin{proof}[Proof of Theorem~\ref{fl:max:thm:main}]
Every decision is a deterministic function of the instance and observed arrival history.
In particular, no auxiliary Poisson sample used in the analysis is generated by the algorithm.
The initialization cost $W_T$ is the cost of the first $\min\{N(T),m\}$ requests, where $N(T)$ is the merged arrival count.
The marks $M_j$ are independent of all merged arrival times by Lemma~\ref{fl:max:lem:observed-seeds}.
Each mark has expected singleton cost $L/\Lambda$, so
$$\E[W_T]=\frac{L}{\Lambda}\sum_{j=1}^m\Pr(A_j\le T)
=\frac{L}{\Lambda}\E[\min\{N(T),m\}]\le\frac{mL}{\Lambda}.$$
There is also the pathwise bound $W_T\le m\max_xs_x$.

The remaining horizon $h=(T-\sigma)_+$ and the assignment $\pi$ can be dependent.
Nonnegativity, rather than an independence assumption, gives
$$\E[h\Psi(\pi)]\le T\E[\Psi(\pi)]\le3T\mu\le3\E[\OPT(I_T)].$$
Average Lemma~\ref{fl:max:lem:scheduled}, use $h\le T$, and enlarge its correction to all facilities.
Adding the initialization bound proves the finite-horizon inequality.

If $L>0$, some positive-rate type has $s_x>0$.
The auxiliary Poisson sample then has positive probability of containing exactly one point, which is accepted.
Hence $\mu>0$ and $\E[\OPT(I_T)]\ge T\mu$ grows linearly.
The initialization, square-root, and constant corrections vanish after division by this denominator, proving the long-run factor $3$.
If $L=0$, all positive-rate types have free singleton service, and the policy costs zero both during and after initialization.
If $m=0$, the empty policy suffices.

Computing the instance parameters, scanning facility sets, sorting at most $m$ candidates, and forming the assignment takes $O(|X||F|+|X|\log|X|)$ operations.
Recording and scaling each initial gap takes constant additional work.
The expected activation time is $\E\sigma=m/\Lambda$, but initialization creates no waiting because every request in that phase is served immediately.
\end{proof}

Fixed finite periods give a deterministic bounded-wait variant.
\begin{corollary}\label{fl:max:cor:bounded-wait}
For $m>0$ and any $\varepsilon>0$, replace the increasing calendar after initialization at every facility with $c_f\Lambda_f>1$ by period $h_f=c_f/\varepsilon$, starting from $\sigma$.
The resulting deterministic policy has maximum wait at most $\max_f c_f/\varepsilon$ and, for every $T>0$,
$$\E[\ALG_{\varepsilon}(I_T)]\le3(1+\varepsilon)\E[\OPT(I_T)]
 +\frac{mL}{\Lambda}+(1+\varepsilon^{-1})\sum_{f\in F}c_f.$$
On instances with positive optimal asymptotic cost, its long-run RoE is at most $3(1+\varepsilon)$.
\end{corollary}
\begin{proof}
Initialization serves immediately and has expected cost at most $mL/\Lambda$.
Conditional on $\mathcal H$, put $h=(T-\sigma)_+$.
A scheduled facility pays at most
$$\left\lceil\frac{h}{h_f}\right\rceil(c_f+h_f)
\le(1+\varepsilon)h+(1+\varepsilon^{-1})c_f.$$
Every such facility contributes one to $\Psi(\pi)$.
Immediate-service and connection costs are unchanged and nonnegative.
Thus the conditional post-initialization cost is at most $(1+\varepsilon)h\Psi(\pi)$ plus the displayed correction summed over scheduled facilities.
Use $h\Psi(\pi)\le T\Psi(\pi)$, average, apply the structural comparison, and enlarge the correction to all facilities.
Adding initialization proves the inequality, and the period bound gives the maximum wait.
\end{proof}

%% file: modules/fl/figures/maximum_intervals.tex
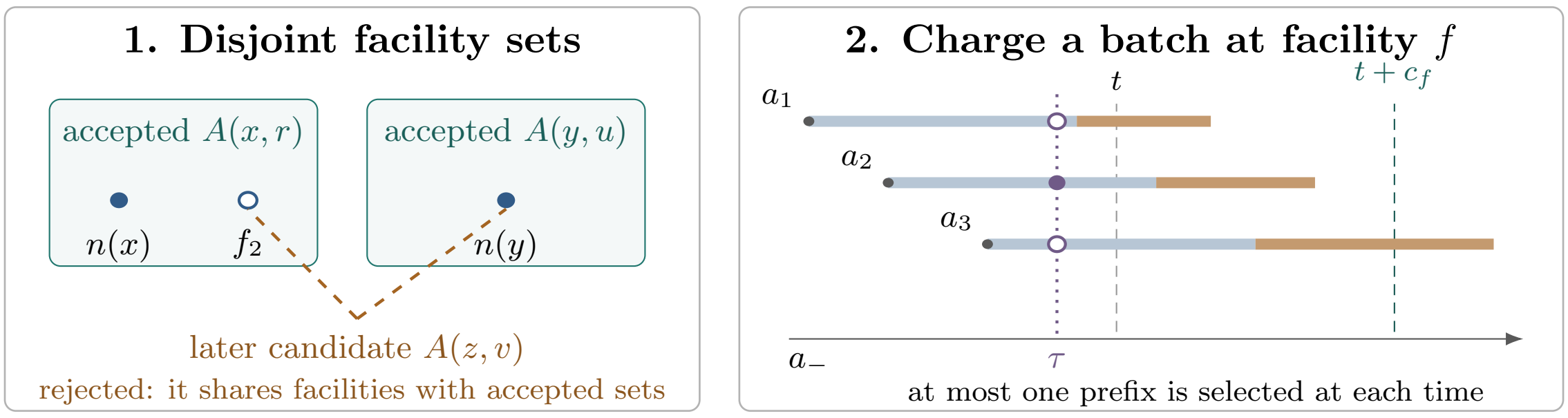
\begin{figure}[t]
\centering
\begin{tikzpicture}[x=.95cm,y=.84cm,>=Latex,font=\footnotesize]
  \tikzset{panel/.style={draw=black!30,rounded corners=3pt,line width=.5pt}}
  \draw[panel] (0,0) rectangle (7.0,4.6);
  \node[font=\bfseries] at (3.5,4.2) {1. Disjoint facility sets};
  \draw[PaperTeal,fill=PaperTeal!5,rounded corners=3pt]
    (.45,1.65) rectangle (3.15,3.55);
  \draw[PaperTeal,fill=PaperTeal!5,rounded corners=3pt]
    (3.65,1.65) rectangle (6.45,3.55);
  \node[PaperTeal!80!black] at (1.8,3.17) {accepted $A(x,r)$};
  \node[PaperTeal!80!black] at (5.05,3.17) {accepted $A(y,u)$};
  \fill[PaperBlue] (1.15,2.4) circle (.09);
  \draw[PaperBlue,line width=.8pt,fill=white] (2.45,2.4) circle (.09);
  \fill[PaperBlue] (5.05,2.4) circle (.09);
  \node[below] at (1.15,2.23) {$n(x)$};
  \node[below] at (2.45,2.23) {$f_2$};
  \node[below] at (5.05,2.23) {$n(y)$};
  \node[PaperAmber!85!black] (later) at (3.55,.7) {later candidate $A(z,v)$};
  \draw[dashed,PaperAmber,line width=.85pt] (later.north)--(2.45,2.30);
  \draw[dashed,PaperAmber,line width=.85pt] (later.north)--(5.05,2.30);
  \node[font=\scriptsize,PaperAmber!85!black] at (3.5,.22) {rejected: it shares facilities with accepted sets};

  \draw[panel] (7.4,0) rectangle (15.7,4.6);
  \node[font=\bfseries] at (11.55,4.2) {2. Charge a batch at facility $f$};
  \draw[-Latex,black!65] (7.9,.82)--(15.3,.82);
  \draw[densely dashed,black!40] (11.2,.9)--(11.2,3.5);
  \draw[densely dashed,PaperTeal!75!black] (14.0,.9)--(14.0,3.5);
  \node[above,font=\scriptsize] at (11.2,3.5) {$t$};
  \node[above,font=\scriptsize,PaperTeal!80!black] at (14.0,3.5) {$t+c_f$};
  \draw[line width=3pt,PaperBlue!35] (8.1,3.3)--(10.8,3.3);
  \draw[line width=3pt,PaperAmber!65] (10.8,3.3)--(12.15,3.3);
  \draw[line width=3pt,PaperBlue!35] (8.9,2.6)--(11.6,2.6);
  \draw[line width=3pt,PaperAmber!65] (11.6,2.6)--(13.2,2.6);
  \draw[line width=3pt,PaperBlue!35] (9.9,1.9)--(12.6,1.9);
  \draw[line width=3pt,PaperAmber!65] (12.6,1.9)--(15.0,1.9);
  \foreach \a/\y/\i in {8.1/3.3/1,8.9/2.6/2,9.9/1.9/3}{
    \fill[black!65] (\a,\y) circle (.055);
    \node[above left] at (\a,\y) {$a_{\i}$};
  }
  \draw[dotted,PaperPurple,line width=.8pt] (10.6,.88)--(10.6,3.62);
  \draw[PaperPurple,line width=.8pt,fill=white] (10.6,3.3) circle (.085);
  \fill[PaperPurple] (10.6,2.6) circle (.085);
  \draw[PaperPurple,line width=.8pt,fill=white] (10.6,1.9) circle (.085);
  \node[below,PaperPurple] at (10.6,.82) {$\tau$};
  \node[below] at (8.1,.82) {$a_-$};
  \node[font=\scriptsize,align=center] at (12.0,.2)
    {at most one prefix is selected at each time};
\end{tikzpicture}
\caption{The common packing rule.
The left panel shows facility incidences, not disjoint geometric balls:
accepted sets share no facility, and filled blue points are selected nearest facilities.
In the right panel, the blue prefix of request $i$ has length
$(s_{x_i}-d(x_i,f))_+$ and its amber tail has length at most $d(x_i,f)$.
All prefixes lie in $[a_-,t+c_f]$.
At any time $\tau$, at most one accepted ball contains $f$, as indicated
by the filled purple point.
The envelope pays the integrated prefix contribution; connections pay the tails.}
\label{fl:fig:maximum-intervals}
\end{figure}

%% file: sections/ai_declaration.tex
\section*{Declaration of AI use}
Generative AI tools (OpenAI's Astra and Codex) assisted mathematical
exploration, proof development, drafting and revision, literature searches,
and LaTeX preparation. The author is responsible for the correctness of
the results and references and for the final text.

%% file: references.tex
\begingroup
\small
\bibliographystyle{plainurl}
\bibliography{references}
\endgroup